\documentclass[12pt,preprint]{emulateapj}
\usepackage{natbib}
\usepackage{graphicx}
\usepackage{amsmath}
\usepackage{xcolor}
\usepackage{url}
\RequirePackage{lineno}

\shorttitle{Recommendation Algorithm to Predict Giant Exoplanet Hosts}
\shortauthors{Hinkel et al.}

\def\gtaprx{ \mathrel{ \vcenter{
      \offinterlineskip \hbox{$>$}
      \kern 0.3ex \hbox{$\sim$}    } } }

\def\ltaprx{ \mathrel{ \vcenter{
      \offinterlineskip \hbox{$<$}
      \kern 0.3ex \hbox{$\sim$}    } } }

\begin{document} 

\title{A Recommendation Algorithm to Predict Giant Exoplanet \\ Host Stars Using Stellar Elemental Abundances}

\author{
  Natalie R. Hinkel\altaffilmark{1,2}, Cayman Unterborn\altaffilmark{3}, Stephen R. Kane\altaffilmark{4},  Garrett Somers\altaffilmark{2}, and Richard Galvez\altaffilmark{5, 6}
  }
\email{natalie.hinkel@gmail.com}
\altaffiltext{1}{Southwest Research Institute, San Antonio, TX 78238, USA}
\altaffiltext{2}{Department of Physics \& Astronomy, Vanderbilt University, Nashville, TN 37235, USA}
\altaffiltext{3}{School of Earth and Space Exploration, Arizona State University, Tempe, AZ 85287, USA}
\altaffiltext{4}{Department of Earth and Planetary Sciences, University of California, Riverside, CA 92521, USA}
\altaffiltext{5}{Center for Data Science, New York University, 60 5th Avenue, 7th Floor, New York, NY 10011, USA}
\altaffiltext{6}{Center for Cosmology and Particle Physics, New York University, 60 5th Avenue, New York, NY 10011, USA}

\begin{abstract}
The presence of certain elements within a star, and by extension its planet, strongly impacts the formation and evolution of the planetary system. The positive correlation between a host starÕs iron-content and the presence of an orbiting giant exoplanet has been confirmed; however, the importance of other elements in predicting giant planet occurrence is less certain despite their central role in shaping internal planetary structure. We designed and applied a machine learning algorithm to the Hypatia Catalog \citep{Hinkel14} to analyze the stellar abundance patterns of known host stars to determine those elements important in identifying potential giant exoplanet host stars. We analyzed a variety of different elements ensembles, namely volatiles, lithophiles, siderophiles, and Fe. We show that the relative abundances of oxygen, carbon, and sodium, in addition to iron, are influential indicators of the presence of a giant planet. We demonstrate the predictive power of our algorithm by analyzing stars with known giant planets and found that they had median 75\% prediction score. We present a list of $\sim$350 stars with no currently discovered planets that have a $\ge$90\% prediction probability likelihood of hosting a giant exoplanet. We investigated archival HARPS data and found significant trends that HIP62345, HIP71803, and HIP10278 host long-period giant planet companions with estimated minimum $M_p\sin(i)$ values of 3.7, 6.8, and 8.5 M$_{J}$, respectively. We anticipate that our findings will revolutionize future target selection, the role that elements play in giant planet formation, and the determination of giant planet interior structure models.

\end{abstract}

\section{Introduction}
\label{intro}

In \citet{Gonzalez:1997p3950}, it was reported that four stars with orbiting giant planets were enriched in [Fe/H]. Since that time, many studies have confirmed that stars with super-solar [Fe/H] are more likely to host giant planets \cite[e.g.][]{Laws:2003p3002,Santos:2004p2996,Bond:2006p2098,Gilli:2006p2191,Sousa:2011p6355}. This trend has been called the ``planet-metallicity correlation," as popularized by \citet{Fischer:2005p948}. In each of these studies, [Fe/H] has been used as a proxy for the star's overall metallicity, that is the total abundance of all elements heavier than H or He. In other words, it has been assumed that the abundance of other elements are consistent with the trends found in Fe relative to H. 

While [Fe/H] is one of the most commonly measured elemental abundances, a variety of different literature sources measured non-Fe elements within giant planet-hosting stars. These elemental ratios were not found to follow follow the [Fe/H] trend, and were not consistent between studies. For example, 
$\alpha$-element enrichment was reported by \citet{Fuhrmann08} and \citet{Adi12a} for giant planet host stars, however, \citet{Mishenina16} found only an overabundance in Al.  Indeed, while some groups measure possible enrichment in certain elements \citep[e.g.][]{Gilli:2006p2191, Robinson:2006p6783,Bond:2006p2098, Brugamyer:2011p3104, Adi12a}, many find that abundances in stars with planets are similar to stars without detected planets \citep[e.g.][]{Fischer:2005p948, Takeda:2007p1531, Neves:2009p1804, delgado10}. \citet{Hogg16} advocated for accurate measurements from a variety of elements since they are important for deciphering the unique chemical signatures between stars while \citet{Hinkel18} argued a similar point but with respect to estimating planetary interior structure and mineralogy. Given this lack of clarity, there is a need to uncover more subtle correlations within stellar abundance data as it relates to planet occurrence that were perhaps hidden from less statistically advanced methods.

For this study, we draw upon machine learning techniques (described in Section \ref{algorithm}) to formulate a method to predict which stars in the solar neighborhood are likely to host to-date undetected giant exoplanets.
Regarding planet detection, there is inherent observation bias (e.g. towards more massive or larger planets) that cannot be overcome at this time. All of the current detection techniques, namely microlensing, radial velocity, transit, and direct imaging, have detection thresholds that only allow a planet to be excluded down to some limit. While some of the main results from the Kepler survey have shown that planets are nearly ubiquitous and that multiplanet systems may be fairly common \citep{Dressing13}, these results are dependent on a relatively small number of observations. Therefore, in order to transform and broaden target selection for exoplanet surveys, it is possible to establish the chemical interplay between planet formation and the composition of the host star.

In this paper we will use the Hypatia Catalog of stellar abundances in order to sample a large set of abundances for stars that do and do not host giant exoplanets. We produce a target list of probable giant planet hosting stars using a variety of element ensembles. In Section \ref{algorithm}, we describe the recommendation algorithm that we employed, the ensembles of elements that we utilized, the Hypatia Catalog, and an overview of the results. We give a list of stars that are likely to host giant exoplanets based on our algorithm in Section \ref{predict}. We discuss our results in Section \ref{discussion}.

\section{The Machine Learning Algorithm}\label{algorithm}
Machine learning is an interdisciplinary field, combining components from statistics, computer science, and pure mathematics, with the aim of extracting meaningful information from existing empirical observations. A ``supervised classifier" concerns itself with the derivation of a generalized relationship that maps existing data (input ``features") from a ``training" set in order to make predictions on new observations (output ``targets"). Here we use decision trees to split a larger set into smaller subsets based on the similarity of properties, in this case whether a star is likely to host a planet or not likely. We use an ensemble of trees, or ``gradient boosted trees," to build a series (as opposed to parallel) set of trees. The trees are trained such that they are able to correct the mistakes of the previous tree in the series, thereby creating a more powerful model for classification. 

An excellent example of a supervised classifier is the movie prediction algorithm used by an online streaming service (e.g. Netflix\footnote{\url{www.netflix.com}}). Namely, after watching a variety of movies (or ``features") on the service, and rating them as liked or disliked (the ``decision tree"), the software is ``trained" to determine the overall relationship between those movies that you like, i.e. they are goofy 90's comedies. It then takes that relationship and searches within its back catalog to find movies that are similar (or ``targets"). Finally, it makes suggestions that the viewer watch movies that are analogous to goofy 90s comedies, with a certain percentage likelihood that they match the overall trend. The decision trees are then improved upon when you watch the recommended movies, thereby building off of (or ``boosting") the previous training models. 

We examine the stellar abundances between stars with and without detected giant planets using the {\it XGBoost} (or Extreme Gradient Boosting) supervised classifier per \citet{Chen16}. From the Hypatia Catalog (Section \ref{sample}), there are 290 planet hosting stars and +4200 stars not currently known to host either giant or rocky planets. Because of the disparity in size between the two populations, we take a random sub-sample of 200 stars from both sets. Then, we train the algorithm to determine the stellar abundance trends for the subset of stars with confirmed giant planets and apply that trend to the subset of stars not known to host planets. The algorithm classifies those stars without detected planets as either likely to host a giant planet (``1") or unlikely to host a giant planet (``0"). This process is run for 3000 iterations choosing a new sub-sample of 200 stars during each iteration or until our model scores no longer changed per iteration of reshuffling. In this way, we are able to produce an overall probability percentage, based on thousands of iterations, that a star not currently known to have a planet is predicted to host a giant exoplanet. The algorithm is available online at: \url{github.com/nhinkel/planetPrediction}. 

\subsection{Sample Selection}
\label{sample}

\begin{figure*}
\begin{center}
 \centerline{\includegraphics[width=19cm]{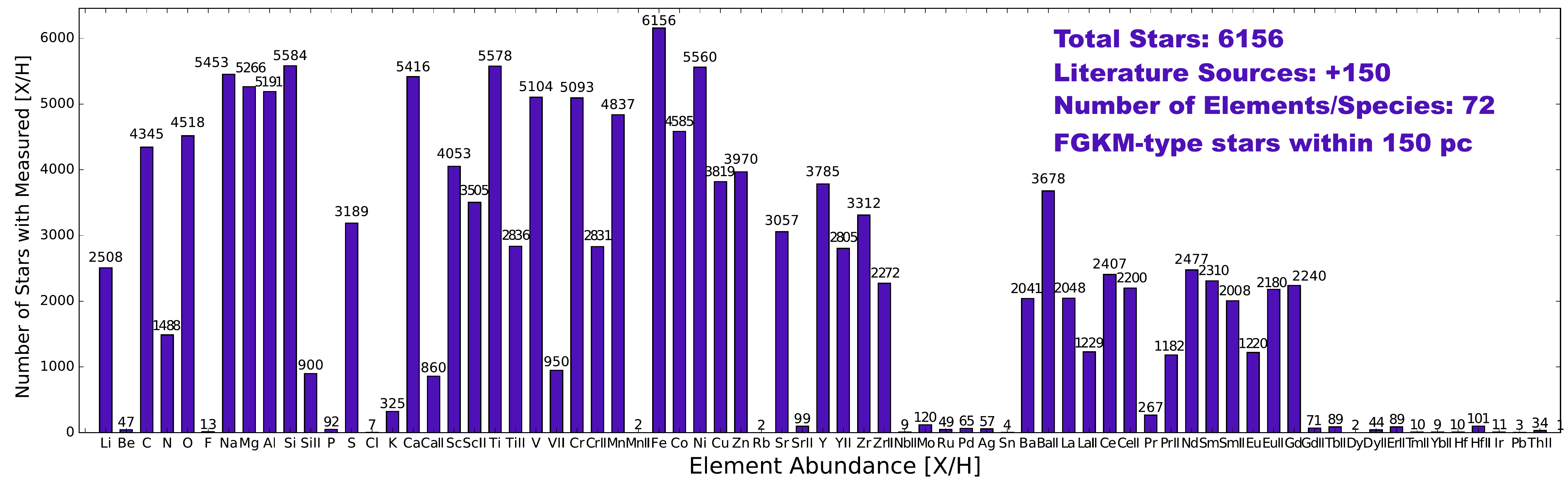}}
  \caption{Number of stars in the Hypatia Catalog that have measurements for each element along the x-axis. The total number of stars in Hypatia is 5986, or the number of iron measurements since [Fe/H] is a requirement to be included in Hypatia. In order to utilize as complete a sample as possible for this study, we analyzed only those elements that were measured within $>$50\% of the Hypatia stars.}
  \label{hist}
 \end{center}
\end{figure*}

The Hypatia Catalog is a database of amalgamate stellar abundance data that currently spans 72 unique elements and species in $\sim$6000 stars within 150 pc of the Sun \citep{Hinkel14, Hinkel16, Hinkel17}, shown\footnote{All data can be found online at \url{www.hypatiacatalog.com}.} in Fig. \ref{hist}. Hypatia is composed of FGKM-type stars, which are ideal for understanding the solar neighborhood, since they are either numerous or intrinsically bright, or both \citep{Freeman02}. 
Hypatia was compiled from +150 literature source abundance measurements -- it is the largest catalog of high resolution stellar abundances for stars within the solar neighborhood. As a result, Hypatia has the breadth (number of stars) and depth (number of elements) that make it uniquely qualified for this analysis. All studies within Hypatia were re-normalized to the same solar scale, namely \citet{Lodders:2009p3091}.  \citet{Hinkel14}, we found that the absolute difference between the individual solar normalizations and the standardized, renormalized abundances varied by 0.06 dex on average and 0.04 dex for the median, which is on par with standard error for many elements. Therefore, because the choice of solar normalization strongly influences the element abundance measurement, we ensure that all values are on a common baseline.
Finally, in those instances where multiple groups measure the same element within the same star, the median value of those measurements was utilized.

In order to focus on stars that have similar compositional trends, we removed all Hypatia stars that likely originated from the thick disk or halo using the kinematic prescription in \citet{Bensby03}. 
Additionally, we require a stellar sample that is as large as possible and also densely populated, i.e. with few null or ``missing" measurements (discussed more in Section \ref{ens-variations}). Therefore, we included only those [X/H] elements that were often measured in nearby stars, or within $>$50\% of the total stellar sample, as shown in Fig. \ref{hist}. Note that the total number of stars with Zn measurements was $<$50\% once probable thick disk stars were removed; also, we chose to use Sc as opposed to Sc II for simplicity within the model \citep{Hinkel14}. 

Within the final sample, 319 stars are known to host giant exoplanets per the NASA Exoplanet Archive\footnote{\url{https://exoplanetarchive.ipac.caltech.edu/} -- more detail regarding this dataset can be found in the Supplement}. To better understand our biases with respect to planet detection, we found that 1 planet was discovered by direct imaging, 2 planets via the transit method, and the remainder were observed via radial velocity. Therefore, we removed those stars that were discovered by any method that wasn't radial velocity. Per \citet{Mayor11, Adibekyan12}, who defined 0.0945 $M_J$ = 30 $M_{\oplus}$ as the mass cutoff between Neptune and Jupiter sized planets, we have excluded any stars with planets below this range. In this way, our total dataset consists of +4200 main sequence stars not known to host planets, 290 confirmed giant exoplanet host stars discovered via the radial velocity method.

\subsection{Ensembles of Elements}\label{ensembles}
One benefit of employing a machine learning algorithm is that it allows us to examine elements within ensembles, instead of on an individual basis. We find that this method of investigation is more meaningful in the context of stars that host planets, since a number of chemical factors must be taken into account during planet formation. Given the large number of elements within the Hypatia Catalog, we wanted to examine a variety of them, and in different combinations, to better understand how the elements may influence each other within the algorithm. We tested our algorithm using the following element combinations or ensembles, where the total number of elements is given in parenthesis at the end of each list:

\begin{description}
\item[Volatiles + Lithophiles + Siderophile + Fe] \hfill \\  C, O, Na, Mg, Al, Si, Ca, Sc, Ti, V, Mn, Y, Cr, Co, Ni, Fe (16) \item[Volatiles + Lithophiles + Siderophile] \hfill \\  C, O, Na, Mg, Al, Si, Ca, Sc, Ti, V, Mn, Y, Cr, Co, Ni (15) \item[Lithophiles + Siderophile + Fe] \hfill \\  Na, Mg, Al, Si, Ca, Sc, Ti, V, Mn, Y, Cr, Co, Ni, Fe (14) \item[Volatiles + Lithophiles + Fe] \hfill \\  C, O, Na, Mg, Al, Si, Ca, Sc, Ti, V, Mn, Y, Fe (13) \item[Lithophiles + Siderophile] \hfill \\ Na, Mg, Al, Si, Ca, Sc, Ti, V, Mn, Y, Cr, Co, Ni (13) \end{description}
\noindent

Our elements were grouped according to a combination of nucleosynthetic origin as well as the preferred host phases for planet building. This was done in order to account for, namely: volatiles (Vol: C and O) which are important for planetary atmospheres; lithophiles that combine to form oxide minerals characteristic of rocks and create part of the core needed for volatile accretion needed to create gas giants (Litho: Na, Mg, Al, Si, Ca, Sc, Ti, V, Mn, Y); and siderophiles that are heavier iron-peak elements which, for Earth-like conditions, alloy with iron and present in planetary cores (Sidero: Cr, Co, Ni). Despite Fe being a siderophile, we include or disclude it individually within the ensembles to see how the other elements were influenced by the presence (or lack thereof) of Fe as a predicting feature. These classifications are important not only for rocky planets, but giant planets as well since these volatile elements create a giant planet's large volatile envelope. Before a giant planet can begin accreting volatile species, it must form a large rock and iron ``core'' containing the lithophile elements and Fe. We explored these ensembles as a way of understanding how the three general categories of elements (volatile, lithophile, and siderophile) influenced one another and the impact when certain key groups were removed from the analysis (such as the volatiles and Fe).
We describe how these five different elemental ensembles influenced the model results in Section \ref{ens-variations}.

\begin{deluxetable}{p{0.7cm}p{0.4cm}p{1.0cm}p{1.2cm}p{0.2cm}|p{0.4cm}p{0.4cm}p{0.1cm}} 
  \tablecaption{\label{predtable}Predicted Giant Exoplanet Host Stars: Vol+Litho+Sidero+Fe}
  \tablehead{
\colhead{HIP} &  \colhead{RA} & \colhead{Dec } & \colhead{Spec} &  \colhead{ V  }  & \multicolumn{3}{|c}{} 
\\
\colhead{ } &  \colhead{(deg)} & \colhead{(deg)} & \colhead{ Type } &   \multicolumn{1}{c|}{  mag } & \colhead{Samp} & \colhead{ Pred} &   \multicolumn{1}{c}{Prob } 
}
62345 & 191.63 & -11.81 & G5V & 6.87 & 2882 & 2882 & 1.0  \\ 
24110 & 77.68 & -44.57 & G8IV/V & 8.71 & 2878 & 2878 & 1.0  \\ 
111978 & 340.23 & -31.99 & K0IV-V & 7.39 & 2856 & 2856 & 1.0  \\ 
20489 & 65.87 & -27.66 & G3/5V & 8.58 & 2863 & 2863 & 1.0 \\ 
81347 & 249.20 & -6.29 & G5V & 7.82 & 2861 & 2861 & 1.0 \\ 
68936 & 211.67 & -5.52 & K1V & 8.36 & 2844 & 2844 & 1.0 \\ 
116823 & 355.17 & 0.42 & K2III & 7.44 & 2846 & 2846 & 1.0 \\ 
71803 & 220.32 & -4.94 & G6V & 8.21 & 2877 & 2877 & 1.0 \\
... & ... & ... & ... & ... & ... & ... \\
\enddata
 \tablenotetext{*}{This is a stub of the table, available via the online journal and Vizier, for the full sample of target stars that were predicted on. The table lists the HIP name, RA, Dec, spectral type, V magnitude, as well as the number of times that the star was sampled (Samp), predicted to host a giant planet (Pred), and the overall probability Pred/Samp) of hosting a giant planet per the Vol+Litho+Sidero+Fe ensemble. For all 5 ensembles of elements, see the table in the Supplement.}
\end{deluxetable}

\subsection{``Golden Sets" and Verification}

The false positive rate, namely where a giant exoplanet is predicted but the presence of one has been observationally excluded, cannot be adequately determined from our analysis. The issue is that while we have star systems with detected or confirmed exoplanets, it is not possible to confirm a null-detection. In other words, we do not know if stars without detected planets are due to physical or chemical reasons or because of observational and/or technological biases. 
For example, nearly face-on orbits of giant planets would be required in order to prevent planetary detection (e.g. a 5$^\circ$ variation from face-on reduces the RV amplitude to $\sim$12\%), however such situations are relatively rare.
Therefore, in order to obtain a quantitative understanding of the accuracy of our algorithm, we created a ``golden set" of known exoplanet planet host stars that was not trained upon, such that it could be predicted upon. We instructed the algorithm to choose 10 random, confirmed exoplanet host stars, the ``golden set," during each iteration of the algorithm and insert them within the sample of target stars, such that the algorithm did not know that they were confirmed planet hosts. We then allowed the algorithm to predict whether the golden set stars were likely to host planets, knowing that they have currently confirmed giant planets. In this way, we are able to test the prediction model to see whether it was able to positively identify stars that we know have planets, thereby testing the ``true positives" without biasing the model. And because we allowed a new set of ``golden" planets to be chosen during each iteration of the algorithm, we were able to analyze thousands of predication probabilities of known planet hosts. 

To illuminate the results of this test we describe here the case of Volatiles + Lithophiles + Siderophile + Fe, which is representative of the results from each ensemble; specific information about the other ensembles can be found in the Supplement. Overall, we find that the algorithm gives a $\sim$75\% prediction score to the known giant exoplanet hosts of our golden samples. This high percentage score indicates these stars are likely to host a giant planet.
As discussed in \citet{Tamayo16}, there is a threshold probability which balances the usefulness of the results (precision) with respect to the completeness of the results (recall or sensitivity).
In other words, if the detection threshold is too low the contamination fraction will be significant, but if the threshold is too high many true positives will be missed. Given our models' significant true positive fraction for all ensembles, we choose to adopt a conservative threshold of 90\%. In this way, we are able to increase the likelihood that a star from our target list with a probability above this value will actually have a yet-undetected giant planet, to be observationally confirmed at a later date. 

To test the robustness of the algorithm, we investigated HARPS radial velocities \citep{Pepe04} for the top 30 stars in Table~\ref{predtable} from the ESO Data Archive\footnote{http://archive.eso.org}. Of the top 30 stars, 20 had publicly available HARPS data, from which the precision radial velocities were extracted via the FITS headers. We then performed a statistical analysis of the data to assess the presence of hot Jupiter companions to the host stars. The semi-amplitude of the radial velocity variations, $K$, due to a companion with mass $M_p$ in a circular orbit with period $P$ may be expressed as:
\begin{equation}
    K = 203.2331 \, \mathrm{m \, s}^{-1} \left( \frac{P}{1 \, \mathrm{day}} \right)^{-\frac{1}{3}} \left( \frac{M_p \sin i}{M_J} \right) \left( \frac{M_\star}{M_\odot} \right)^{-\frac{2}{3}} \,\, ,
    \label{kamp}
\end{equation}
where $i$ is the inclination of the orbit, $M_J$ is the mass of Jupiter, and $M_\star$ is the mass of the host star. Edge-on orbit semi-amplitudes for hot Jupiters can thus range from $\sim$2~km\,s$^{-1}$ for a 10~$M_J$ planet in a 1 day orbit, to $\sim$50~m\,s$^{-1}$ for a 0.5~$M_J$ planet in a 10 day orbit. Our analysis of the 20 stars with radial velocity data found no evidence of planetary signatures with periods less than the time baseline of the data, and so we are able to rule out hot Jupiter companions for these stars. A caveat to note is that, as seen in Equation~\ref{kamp}, the semi-amplitude of the variations are sensitive to the inclination of the orbit. Thus, these stars could also harbor planetary companions in near face-on orbits whose signatures are too small to be detected by the HARPS data. We further performed a linear least-squares fit of the data to assess the significance of trends in the radial velocities that may be indicative of long-period giant planet companions to the stars. This analysis revealed significant trends in the data (correlation coefficient larger than 95\%) for HIP~62345, HIP~71803, and HIP~10278, with trends of 0.025, -0.029, and -0.104 m\,s$^{-1}$\,day$^{-1}$ , respectively. There is no sign of curvature in the observed trends, and Equation~\ref{kamp} was used to estimate minimum $M_p \sin i$ values of 3.7, 6.8, and 8.5 $M_J$ for HIP~62345, HIP~71803, and HIP~10278, respectively.

\subsection{Algorithm Results}
\label{results}
\begin{figure*}[h]
\begin{center}$
\begin{array}{lll}
\includegraphics[trim=0 0 20mm 0,clip,width=.4\textwidth]{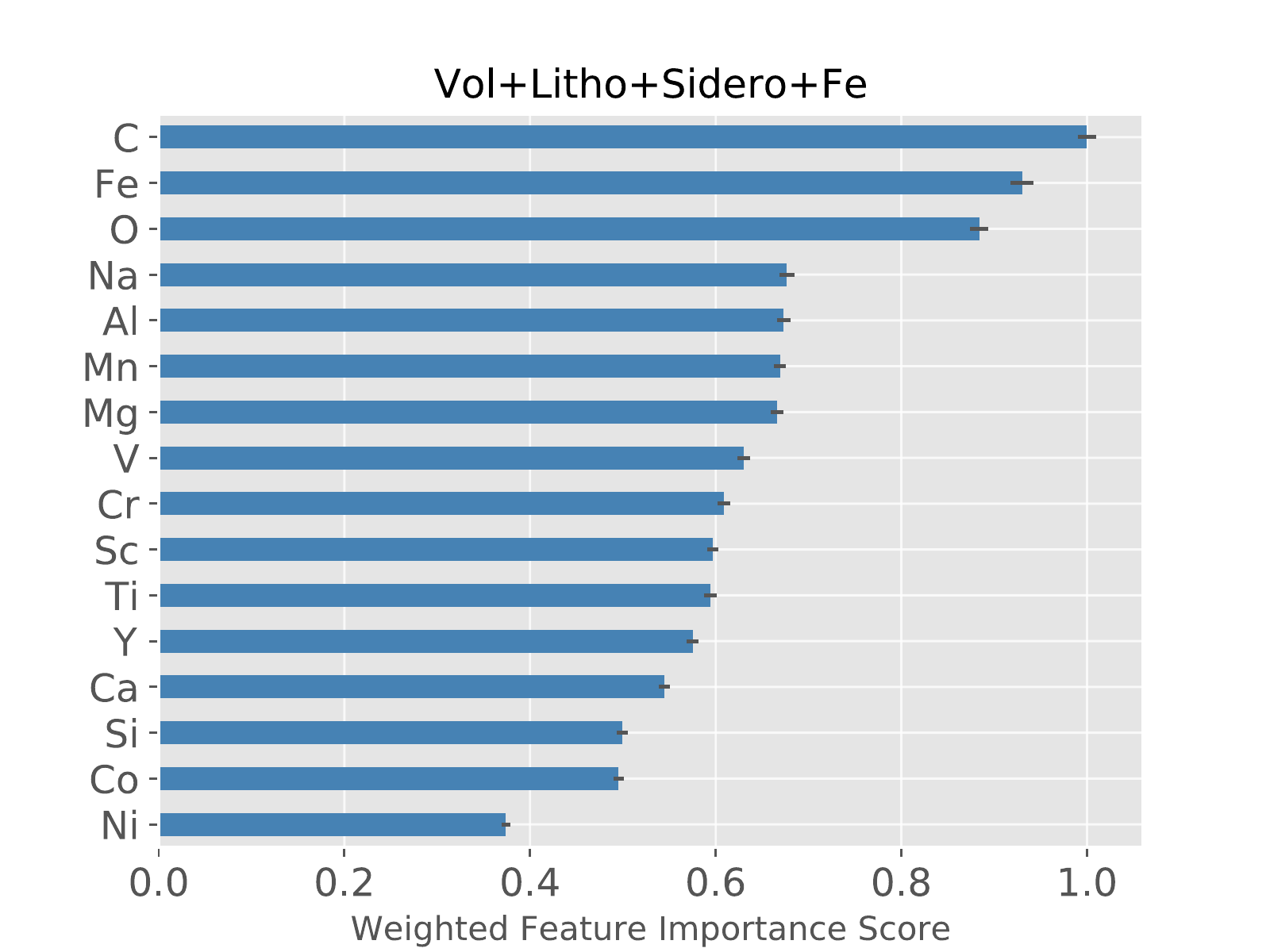}&
\includegraphics[trim=0 0 20mm 0,clip,width=.4\textwidth]{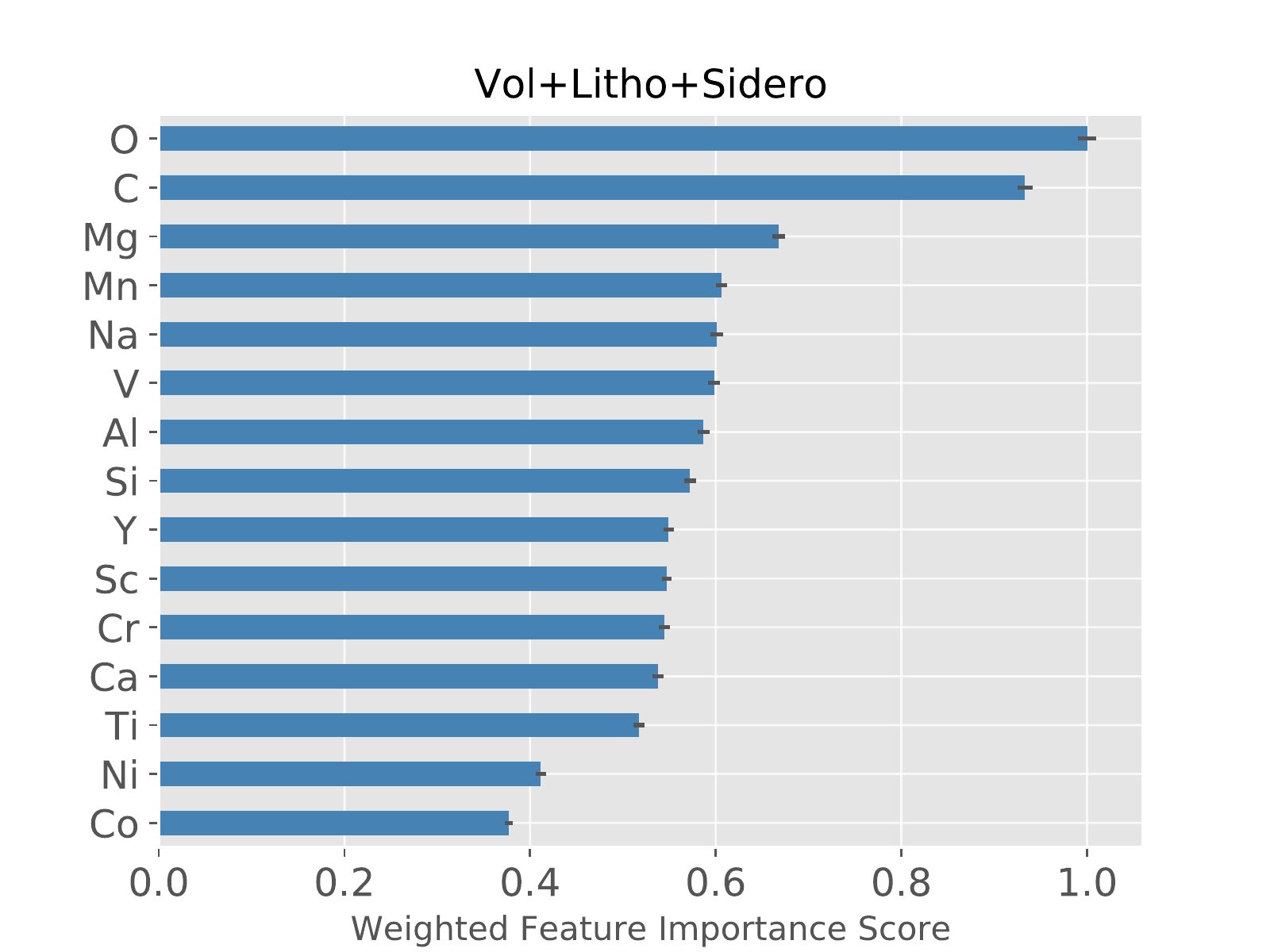}
\end{array}$
\end{center}
\begin{center}$
\begin{array}{rr}
\includegraphics[trim=0 0 20mm 0,clip,width=.4\textwidth]{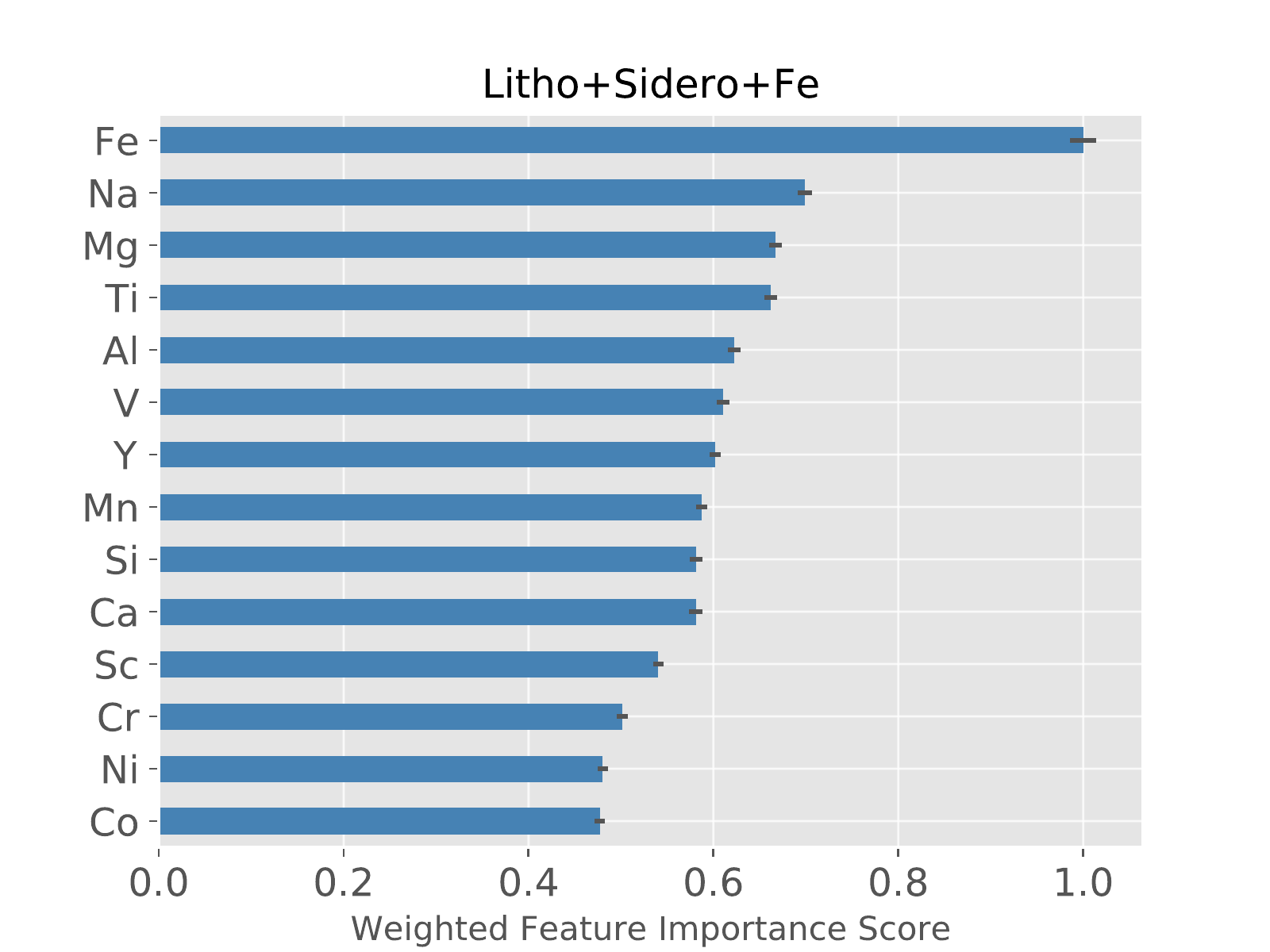} &
\includegraphics[trim=0 0 20mm 0,clip,width=.4\textwidth]{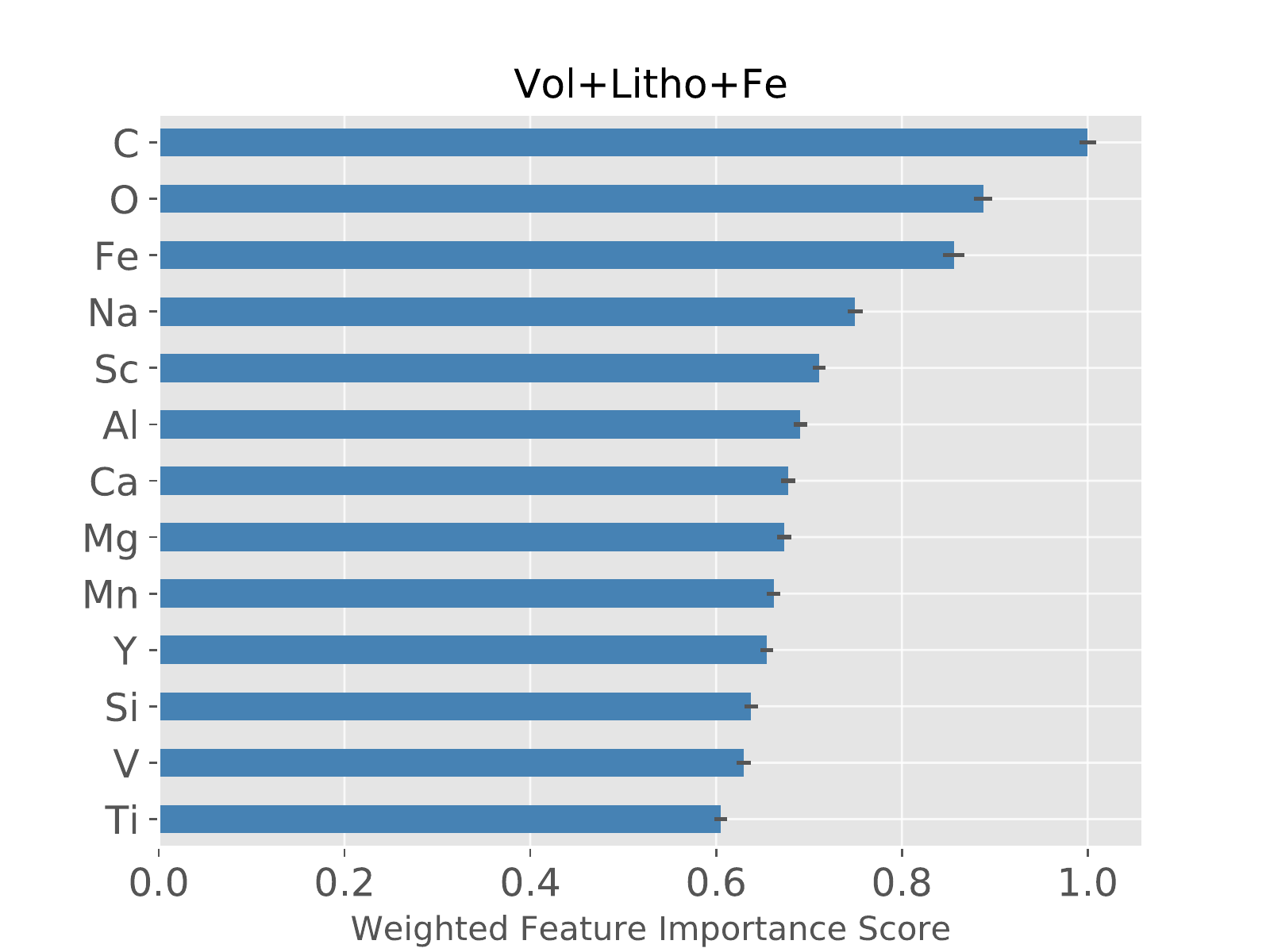}
\end{array}$
\end{center}
\begin{center}$
\begin{array}{rr}
\includegraphics[trim=0 0 20mm 0,clip,width=.4\textwidth]{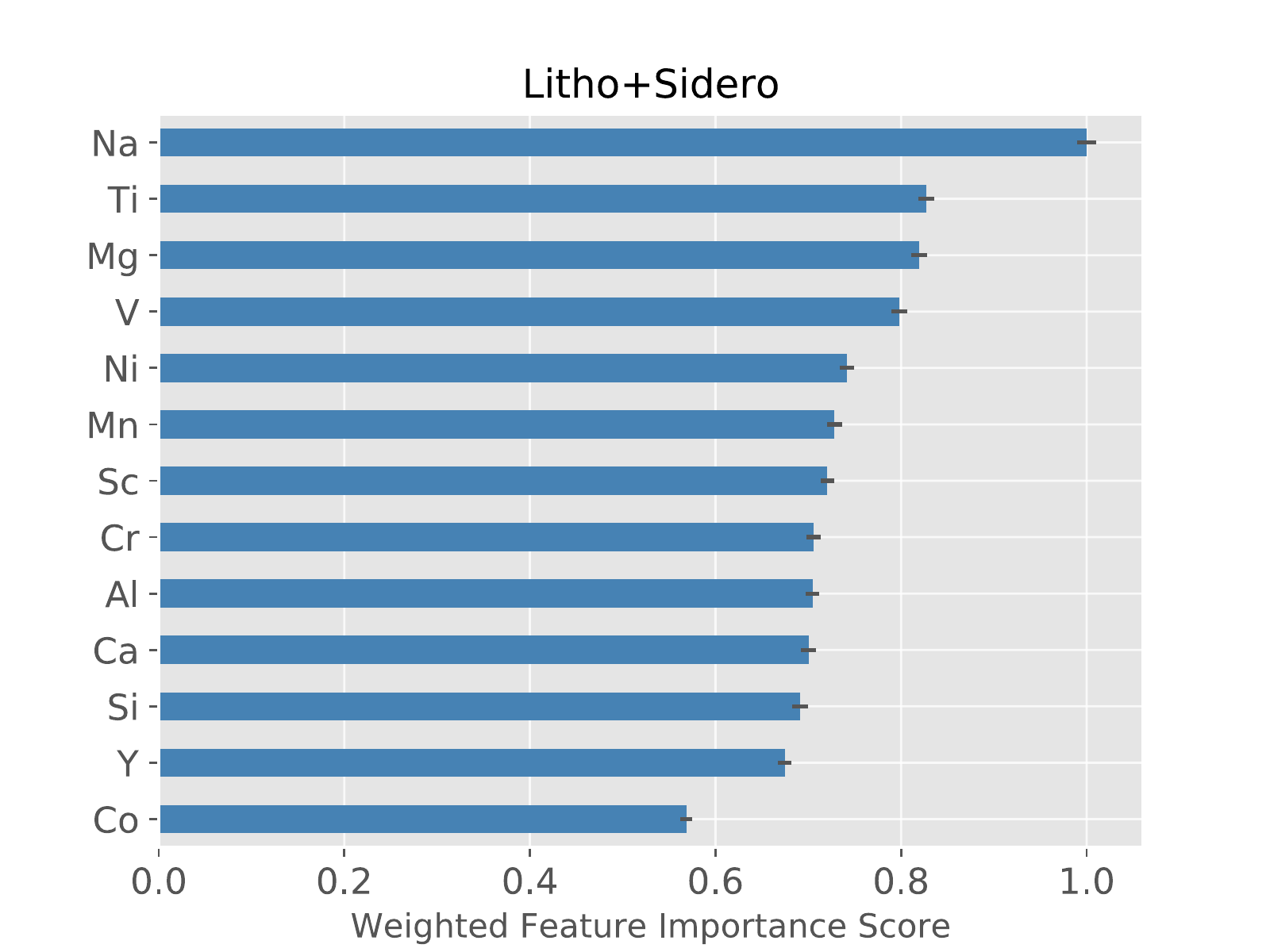}
\end{array}$
\end{center}
\caption{The weighted feature importance scores for all of the ensembles of elements, indicated by the titles on each subfigure. The error bars were calculated by determining the standard error of the mean, by dividing the standard deviation of the abundance measurements by the square root of the number of abundance observations.}
\label{fig:importance}
\end{figure*}

The important input ``features" are those elements that are influential in splitting the overall sample into stars likely to host a giant exoplanet or not (see the Supplement for more information). The weighted feature importance scores for the five ensembles are shown in Figure \ref{fig:importance} -- where the titles at the top of each subfigure indicate the ensemble being represented. The scores are ranked such that the higher the feature importance score, the more influential the element is in lowering the entropy of the dataset. Error bars are given on the right edge of each bar in the horizontal histograms, which were calculated by taking the standard error of the mean. Moving from those ensembles with the most number of elements to the fewest (namely, from top to bottom of Figure \ref{fig:importance}), we see that, when present, the volatile elements C and O, along with Fe, are ranked highest as the most important features for determining whether a star is likely to host a giant planet. When the volatile elements are not within the ensemble (Litho+Sidero+Fe and Litho+Sidero), Na is the most important feature. Additionally, Na is consistently clustered near the top of the ranked features when C and O are present. The only instance where Na is not ranked directly below C, O, and Fe (when present) is during the Vol+Litho+Sidero ensemble, when it is supplanted by both Mg and Mn. To a lesser extent, Al and Mg are typically ranked at a medium to medium-high importance, although, their variation in importance between ensembles precludes a clear interpretation of their overall impact. More details regarding the feature importance scores can be found in the Supplement.

The abundance data within the Hypatia Catalog draws from hundreds of sources, each of which uses their own stellar atmosphere models, spectrum synthesis code, spectral-atomic line list, calibration of stellar parameters and instrumentation. Even for measurements of the same star, these systematic differences in methodology will consequently produce different abundance determinations of individual elements. While correcting for each of these methodological choices and creating a true apples-to-apples sample for comparison would be ideal, this is currently infeasible given that is still not understood the nature of these variations \citep{Smiljanic14, Jofre15, Hinkel16, Jofre18}. In order to verify that our findings were robust with respect to the uncertainty typically associated with the element abundances and not the result of data compilation, we created sub-datasets that vary [X/H] abundances in a Monte-Carlo fashion to within each element's uncertainty in the Hypatia Catalog \citep{Hinkel14}. We performed this random variation five unique times, rerunning the algorithm each time using the Vol+Litho+Sidero+Fe ensemble -- to see how these changes impacted all of the elements. The resulting feature importance scores fluctuated somewhat, but were consistent to within the standard error of the mean. For all five varied abundance sets, the top importance features were C, O, Fe, and to a lesser extent Na. Additionally, the ``golden set" prediction score and the list of predicted giant planet host stars did not significantly vary for the five sub-datasets.

\begin{figure*}
\begin{center}$
\begin{array}{lll}
\includegraphics[trim=0 1mm 0 0,clip,width=.5\textwidth]{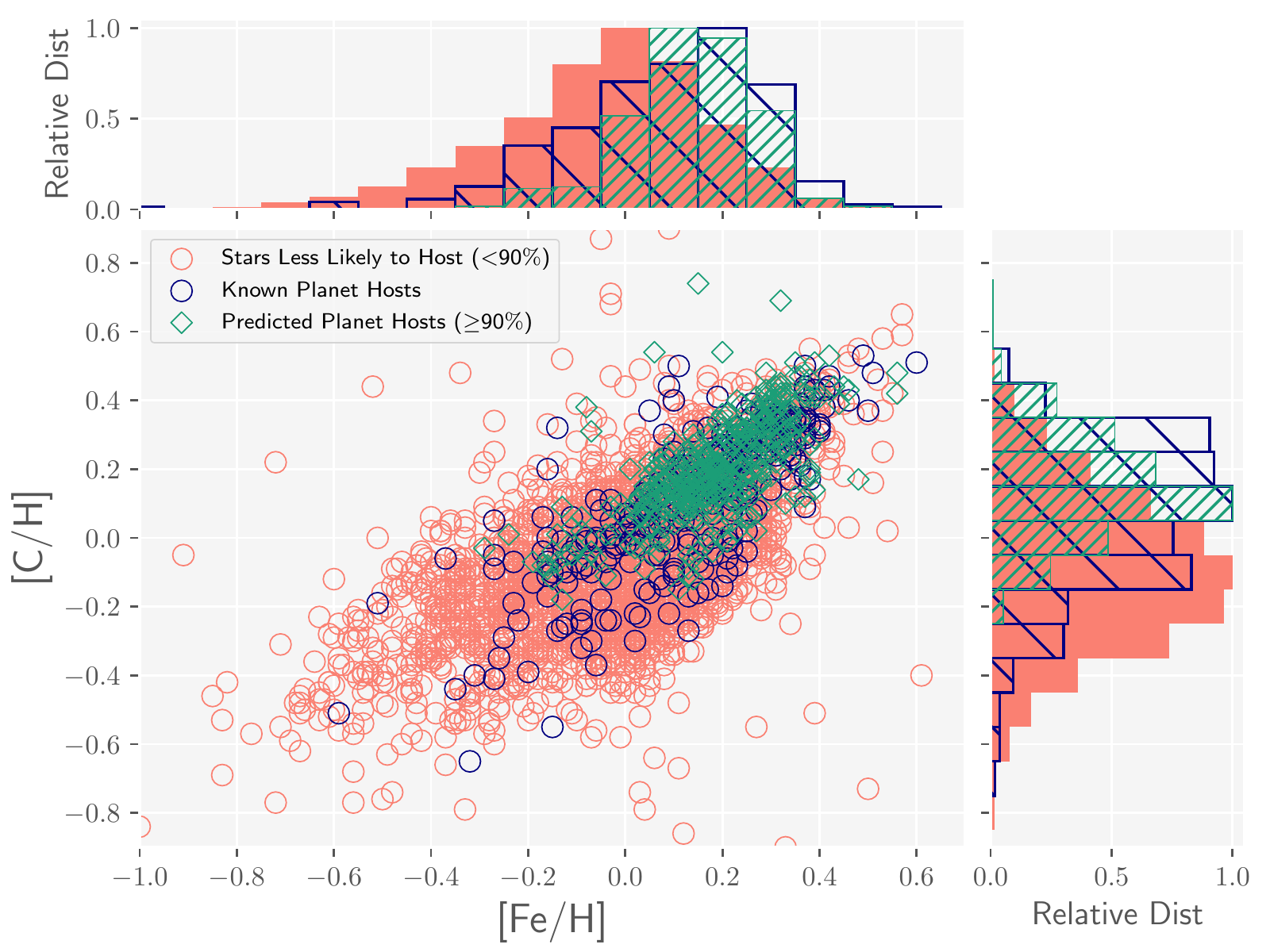}&
\includegraphics[trim=0 1mm 0 0,clip,width=.5\textwidth]{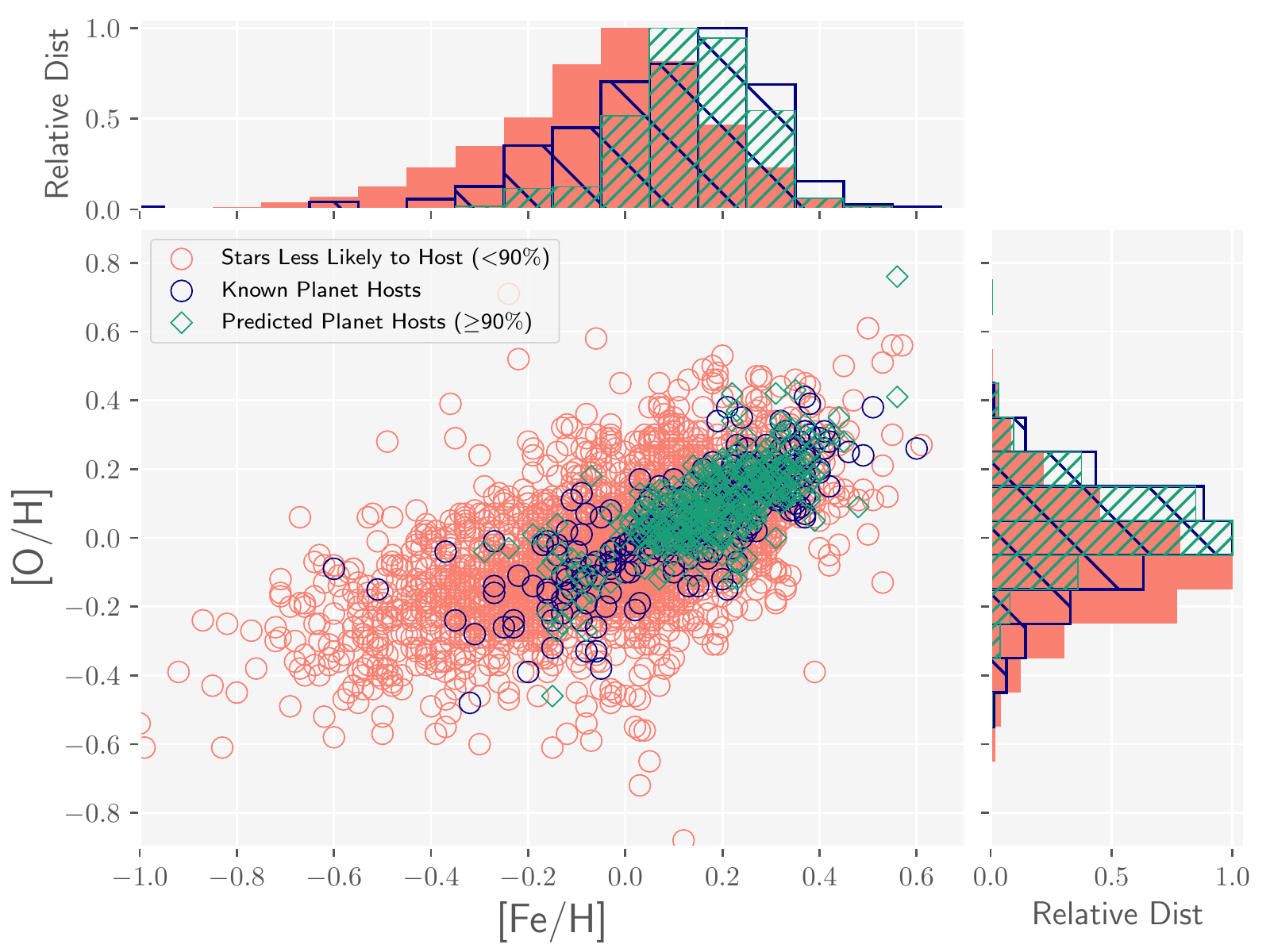}
\end{array}$
\end{center}
\caption{Multiple [X/H] vs [Fe/H] plots for C and O where groups of stars are color-coded to indicate stars less likely to host giant planets (orange), known planet host stars (navy), and stars that were given a $\ge$90\% prediction score via the algorithm (green). One-dimensional histograms of the relative [X/H] distribution are shown to the right of all scatterplots while the [Fe/H] is shown at the top. The bins of all histograms have a width of 0.1 dex.}
\label{fig:CO}
\end{figure*}

As a further verification of the algorithm, we have provided the individual [X/H] vs [Fe/H] plots for C and O in Figure \ref{fig:CO}. The training sample of known planet host stars is given in navy while the target sample of stars not know to host planets are given in either green if they had a $\ge$90\% probability of hosting a giant planet or orange if they were less likely to host a giant planet ($<$90\%). Histograms are provided on each axis in order to verify that similarity between the navy and green samples of stars. More discussion can be found in Section \ref{predict} and in the Supplement.

\section{Predicted Potential Giant Exoplanet Hosts}
\label{predict} 
Using our recommendation algorithm, we have compiled a table of the +4200 target stars with the predicted probabilities that these stars host a giant planet as determined from the Vol+Litho+Sidero+Fe ensemble (Table \ref{predtable})\footnote{Full table can be found via the online journal or Vizier with, an extended version for all ensembles can be found in the Supplement}. 
Probabilities are determined such that the number of times it is positively predicted to host a giant planet (Pred) is divided by the number of times each star was sampled (Samp), such that Prob = Pred/Samp. 
For the Vol+Litho+Sidero+Fe ensemble, 368 stars, or $\sim$9\% of the Hypatia stars we predicted on, have prediction probabilities of hosting a giant planet $\ge$90\%. None of these 368 stars have yet had any discovered planets and thus we  include the RA/Dec, spectral type, and V magnitude in anticipation of potential future observations to detect giant exoplanets orbiting these stars (Table \ref{predtable}). 

In order to fully interpret the results of Table \ref{predtable}, we must understand the biases within the data. Namely, all of the known giant exoplanet host stars that were used as a training set were detected by today's standards (and uncertainties) using the radial velocity method. Therefore, the predicted planet hosting stars are going to have similar detectability biases toward the radial velocity methods. The elements that were used as features for the model are elements that are often measured within stars because they have a relatively large number of clean, unblended lines in the optical band.  Additionally, all of the stars that were given a $\ge$90\% probability prediction were within 100 pc of the Sun, which may be correlated with the fact that it is easier to measure high resolution stellar abundances for nearby stars. Overall, it is our hope and intention to observe the stars with a high probability of hosting a giant planet. However, we recognize that there are caveats within our algorithm, such that a null detection of a giant planet constitutes a reflection of the data for the currently known planet hosts and their abundances, as opposed to the algorithm and generated models. This agnostic, machine learning-based approach shows, however, the power of analyzing multidimensional datasets to elucidate more subtle trends that must be explained by and incorporated into theoretical models of planet formation.

Since there is a tendency for smaller stars to not host giant planets \citep[e.g.][]{Johnson10}, we look at the spectral type distribution for the stars with a high probability ($\ge$90\%) of hosting a giant planet.\footnote{Other physical and chemical properties can be analyzed by the interested reader at \url{www.hypatiacatalog.com}, where the HIP identifiers can be copied directly into the ``Stellar Data Table."} We found that 37 stars (10\%) were F-types, where 1 was a giant and 5 were subgiants. A total of 225 stars (62\%) were G-types, such that 9 were giants and 37 were subgiants. The remaining 104 stars (28\%) were K-types, where 46 were giants and 13 were subgiants. As a comparison, the full sample of target stars contained 22\% F-types, 50\% G-types, and 28\% K-type stars. Therefore, a first-order analysis reveals that the models generated by the algorithm have a modest preference towards G-type stars as giant planet hosts while moving away from F-type stars as giant planet hosts. This trend is likely the result of the fact that F-type stars are faster rotators with higher temperatures, which make precision abundance determinations difficult. As a result, C and O abundance measurements (which are already notoriously difficult to calculate, see \citealt{Ecuvillon2005, Nissen14}) were not successfully determined for as many F-type stars as G- and K-type stars. Given that C and O were both highly ranked importance features, this meant that the algorithm did not find F-type stars to be have high planet prediction likelihoods.

\begin{figure*}[t!]
\begin{center}
 \centerline{\includegraphics[width=0.8\linewidth]{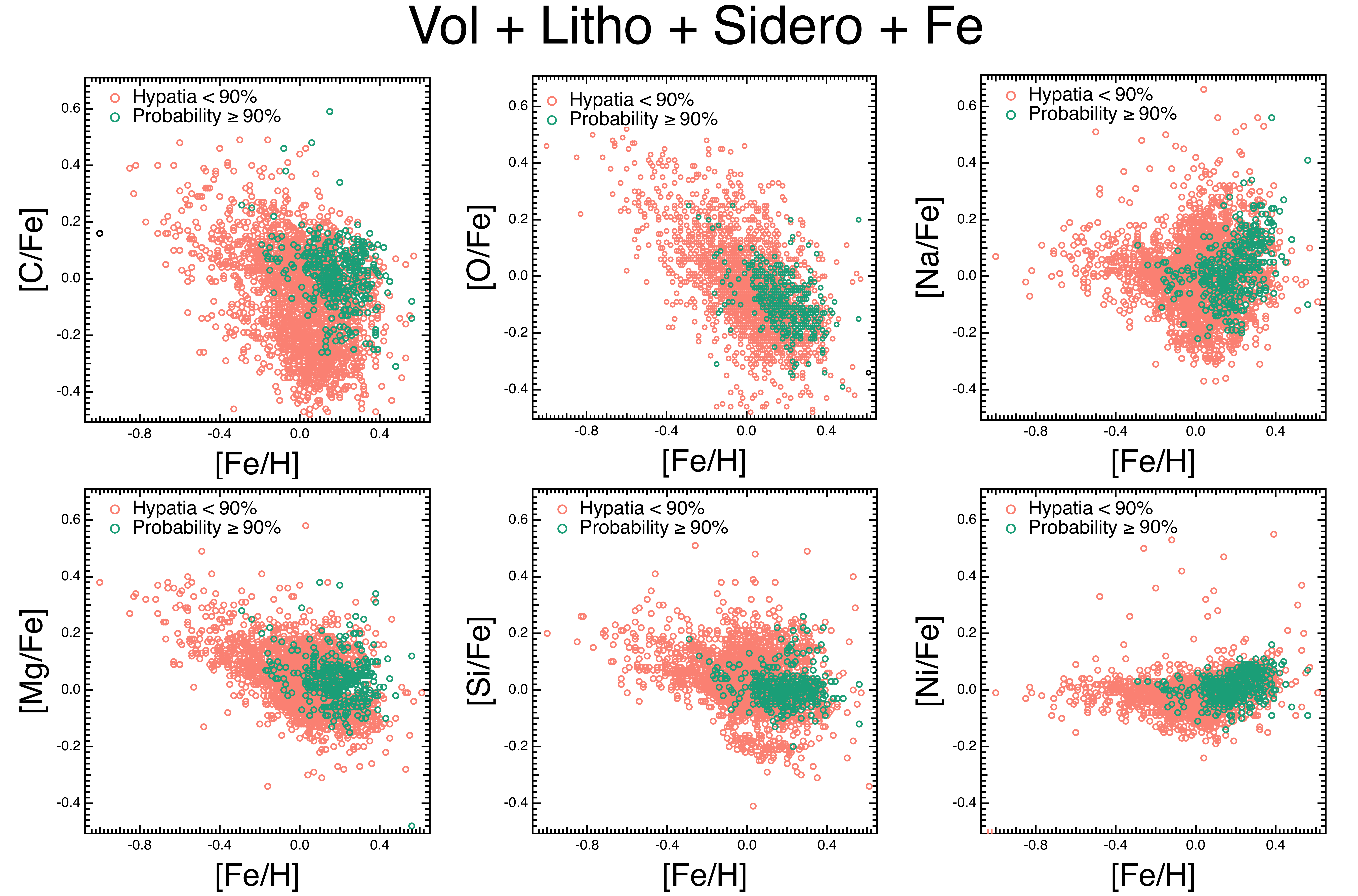}}
\end{center}
  \caption{[X/Fe] vs [Fe/H] for C, O, Na, Mg, Si and Ni for the Vol+Sidero+Litho+Fe ensemble. Those stars with a probability to host giant planets greater than 90\% are shown in green and the remaining stars in the Hypatia sample with a $<$90\% prediction probability are shown in orange, similar to Figure \ref{fig:CO}. Note that all stars in the Hypatia sample are not believed to originate from the thick disk, per the \citet{Bensby03} thin-to-thick disk probability ratios and a more specific chemical analysis for those stars with low-[Fe/H] and enriched [$\alpha$/Fe].}
  \label{fig:CONaProbs}
\end{figure*}

\section{Discussion}
\label{discussion}
\begin{figure*}[t!]
  \centerline{\includegraphics[width=.5\linewidth]{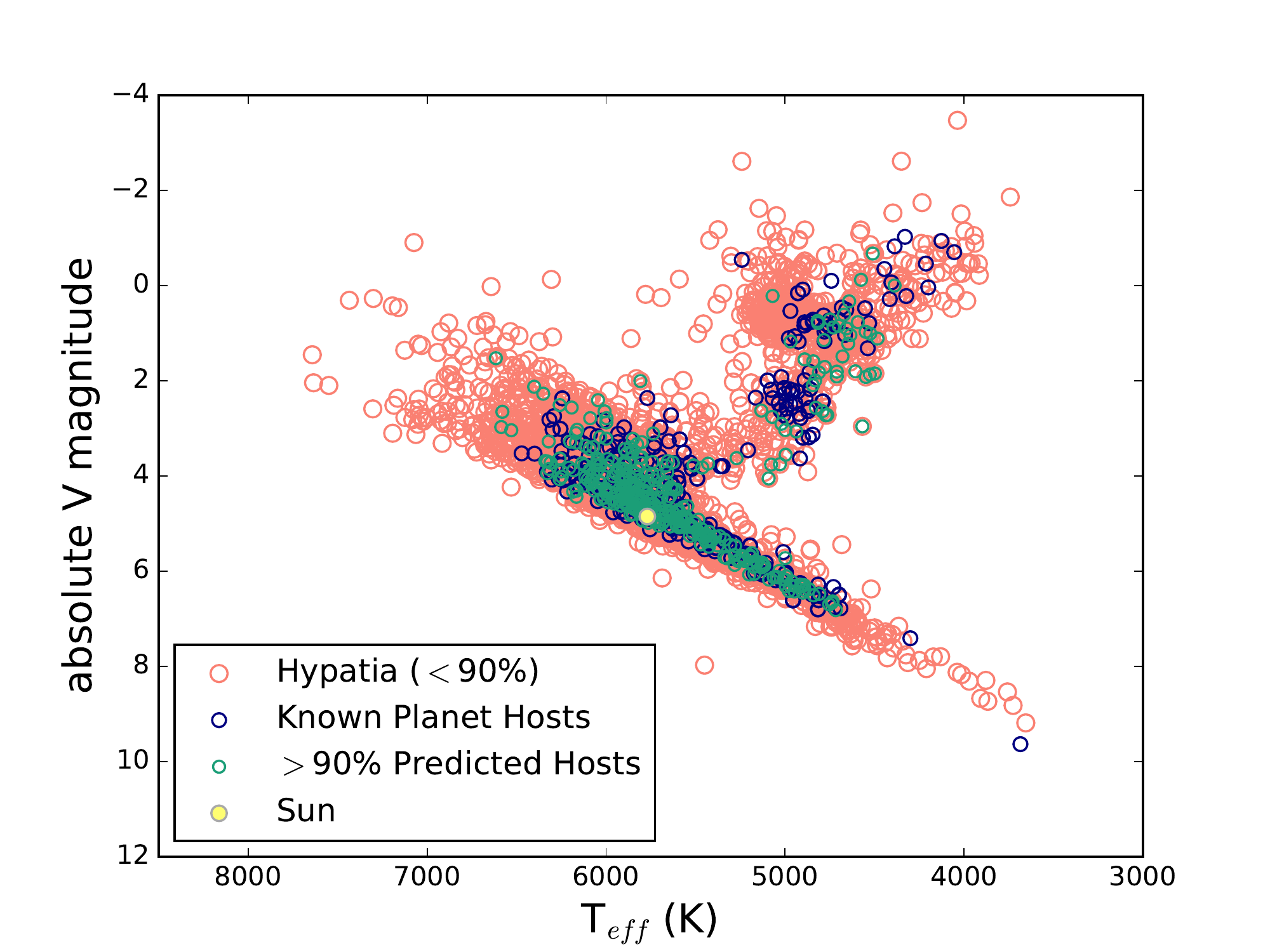}}
  \caption{Hertzsprung-Russell diagram with the effective temperature on the x-axis and absolute V magnitude on the y-axis. All of the stars with a $>$90\% probability of hosting a giant exoplanet are shown in light green, those with $<$90\% probability within Hypatia are given in dark green, and the known giant exoplanet host stars are in gray open circles. The Sun is shown in yellow for comparison. 
 }\label{coprob}
 \vspace{8mm}
\end{figure*}

We analyze the [X/Fe] abundance gradients of stars predicted to host a giant exoplanet within the Vol + Litho + Sidero + Fe ensemble as a function of [Fe/H] (Fig. \ref{fig:CONaProbs}). By analyzing the distributions in this way, we are able to understand the relationship between [Fe/H] and individual element abundances relative to Fe specifically with respect to those star with a planet prediction probability $\ge$90\% and the Hypatia stars with $<$90\%. It it noticeable that the stars with $\ge$90\% prediction probability do not consistently emulate the trends of the full sample. In the case of C, the $\ge$90\% stars are not only enriched in [Fe/H] but also in [C/Fe].  This means that our predicted giant planet hosts have super-solar abundance in C and Fe. Both Na and O show similar trends. [Mg/Fe] and [Si/Fe] for those stars with  prediction probabilities  $\ge$90\%, however, do not show a strong correlation with iron above [Fe/H] $>$ 0 dex, despite the full Hypatia sample slightly negative slope with [X/Fe] vs [Fe/H]. Ultimately, the variations between the two populations illustrate how our algorithm takes into account a combination of the elemental trends within the ensembles, especially as they relate to the known planet host stars, which were the training set. As a result, we are not seeing a simple pattern where most of the stars with [Fe/H] $>$ 0.0 dex are likely to host a giant planet. Instead, the results from the algorithm are more nuanced and hint at other explanations including possible relationships with condensation temperature, planetary dynamics, nucleosynthetic origins, or stellar age. We do find that the majority of those stars predicted to host a giant planet are on the main sequence (Fig \ref{coprob}). However, while these trends are interesting, the interpretations of them are beyond the scope of this paper and are left for a more nuanced future analysis that can analyze these trends with respect to the star-planet interaction.

It is apparent from the major giant planet formation theory, namely core-accretion \citep{Pollack:1996p7814, Ida:2004p8023, mordasini09}, that heavy elements are a fundamental component to the formation process. And while there is clear, empirical evidence of the influence of Fe, it hasn't been established that Fe is more strongly correlated with giant planet occurrence than any of the other elements---even though it theoretically follows that other elements would be necessary for forming massive gaseous planets. As pointed out in \citet{Adi12a}, a pressing problem in the stellar abundance field is that many of the stellar abundance studies are limited to small samples of stars with and without planets. These analyses of small-number statistics can often be contradictory, such that the broader distribution of stars and planets within our solar neighborhood is very difficult to understand. It is with this in mind that we set out to explore the role of stellar composition, especially for non-Fe elements, and the presence of an orbiting giant exoplanet using a novel technique. 

We have utilized an inverse, data-driven approach in order to recommend which solar neighborhood stars may be hosting yet-undetected giant planets based on their stellar elemental abundances. We used the stars and abundances within the Hypatia Catalog \citep{Hinkel14} as a way to achieve a large number of features (elements) and targets (stars that aren't known to host planets). We chose a supervised classifier, specifically the XGBoost algorithm \citep{Chen16}, that would allow us to train the models on the stellar abundances of stars with confirmed giant exoplanets, as determined by the NASA Exoplanet Archive. We utilized the algorithm to predict the likelihood that +4200 FGK-type stars host a giant exoplanet, implementing five different ensembles of elements composed of volatiles, lithophiles, siderophiles, and Fe. Between the ensembles we found that C, O, and Fe, as well as Na although to a lesser extent, are the most important features for predicting giant exoplanet host stars.

Given that C, O, and Fe were the most important features within the latter three ensembles, we wanted to better understand their relationship and influence on one another. Therefore, we ran the algorithm using only those three elements within the ensemble. The resulting feature importance scores yielded Fe as definitively more important (maximum score = 1.0) while C had a weighted score of 0.53 and O had a score of 0.39. This implies that the presence of the other elements within the ensembles (e.g. the lithophiles and siderophiles) had an impact on how the decision tree behaved with respect to both C and O. However, we also see a significant overlap of stars with a high probability ($\ge$90\%) between both the Vol+Litho+Sidero+Fe and Vol+Litho+Sidero ensembles (the probability results and comparison between the ensembles can be found in the Supplement). We interpret this to mean that, while the exact ordering of C, O, and Fe may be influenced by other elements, the three elements (as well as Na) are notably significant in predicting whether a star is likely to host a giant planet.

We note that work done by \citet{Fischer14} and \citet{Howard16} on the Lick and Keck Planet Searches deemed $\sim$30 stars unlikely to host giant planets. However, the Lick and Keck Planet radial-velocity searches were inherently biased towards short-period planets, such that systematics and zero-point offsets may have compromised confirmation of long-period planets \citep{Fischer14,Howard16}. Namely, there could be much longer period planets beneath the detection threshold, and since the orbital inclination is a free parameter, it is possible that these stars could have giant planets in near face-on orbits. Because of these caveats, we have opted not to remove the $\sim$30 stars from our sample. Nevertheless, while the current limits of giant planet detectability can be best described as a gradient (rather than a binary), we recognize that their work is a significant step towards actual ``true negatives."

Similarly, it is possible that the machine learning algorithm presented here may predict the presence of a planet which cannot be confirmed, or a ``false positive." The question then arises as to whether there is a specific incompatibility between the observations and abundances or whether we can draw conclusions on the dynamical history of the planetary system. In the latter case, significant interactions usually occur early in the history of a system which can lead to truncation of planet formation in some radial regions around the star or involve close encounters that result in the planet ejection \citep{Guillochon11, Cloutier15}. Thus, if a system is expected to harbor a planet based on the stellar abundance measurements, but no evidence of such a planet is found, it could indicate a planet in a near face-on orbit or that a planet formed but was subsequently removed via dynamical effects.

Finally, in order to test the accuracy of our recommendation algorithm, and without the availability of stars that are definitively without an orbiting giant planet or ``true negative" cases, we implemented a ``golden set" of stars. Namely, we segregated a group of stars with planets such that the models were not trained using their properties and then ``hid" those stars in the target sample. We allowed the algorithm to predict on the ``golden set" of stars and found that they had an average of $\sim$75\% probability of hosting a giant exoplanet, where more than half of the ``golden set" had a prediction probability $\ge$90\%. We conclude that those stars with a high prediction probability are therefore likely to host a giant planet.

The majority of stellar abundance analyses, specifically as they relate to stars with and without planets, examine the two populations on a single element-by-element basis. However, the benefit of the algorithm developed here is that incorporates information from all elements as well as known planet hosting stars when predicting which stars are likely to host planets. Namely, the population of stars with a $\ge$90\% chance of hosting a planet were identified not just because they were enriched in particular elements, such as Fe, but because the confluence of this particular element with others (C, O, and Na) was unique. It is for this reason that we do not see comparable trends in elements with similar nucleosynthetic origins such as Mg and Si, which are both $\alpha$-elements, or those elements that are important for rocky planet building. Instead, we observe a variation in the [X/Fe] vs. [Fe/H] slopes between the populations of stars that are likely and unlikely to host planets which seem to fold in information that describes the interplay between the star and its planet. The algorithm presented here presents a relatively unbiased method to look for abundance trends within a variety of stars and determine which elements are the most strongly correlated with the presence of planets. Indeed, we argue that it is the data that is potentially biased, not the predictive algorithm.

It is not clear whether there is a correlation between stellar composition and the presence of rocky planets \citep{Buchhave14, Buchhave15, Wang2015, Petigura18}. However, this data-driven and ensemble-based approach may help elucidate any relationship between disk chemistry and the likelihood of a star's ability to host smaller rocky planets, whenever a statistically significant number of nearby, Earth-sized planets has been observed. Indeed, measurement of stellar abundances, \textit{in general}, can greatly aid in our understanding of rocky planet composition \citep[e.g. Mg, Si, Fe: ][]{Dorn15,Unterborn16}, interior structure \citep[ e.g. Mg, Si, Fe, Al, Ca: ][]{Hinkel18}, dynamical state \citep[e.g. C, Th: ][]{Unterborn14, Unterborn15}, the ability sustain detectable biospheres (e.g. P). Many of these ``interdisciplinary elements'' are often overlooked in spectroscopic studies, because their lines are weak, blended, or not readily found within the optical band. However, it is vital to our comprehensive, interdisciplinary understanding of planet diversity and potential habitability that these elements be incorporated into future observations \citep{NAS18}.

While we have thoroughly vetted the recommendation algorithm for biases and systematics, it cannot be denied that the data on which we are training is significantly biased. Namely, we are partial towards radial velocity detected giant planets that are orbiting close to nearby stars with easy to measure elemental abundances. However, upcoming missions like TESS and CHEOPS will be observing a significant number of giant and rocky exoplanetary systems, which will will require ground-based follow-up observations for planet confirmation. It is our hope that the list of predicted giant planet host stars presented here will be used in tandem with the objects of interest from these missions, if for no other reason than a more complete characterization of the system. Indeed, if each of RV follow-up studies include the host star element abundances for those systems found to have planets, predictive models such as this can only be improved. In the future, when more rocky exoplanets have been discovered, we plan to apply our algorithm to the abundances of the host stars. In this way we seek to better understand the trends, if they exist, between the various stellar elemental abundances, both individually and as an ensemble, and the presence of rocky planets.

\section*{Acknowledgements}
The authors would like to thank Megan Bedell, Josh Kammer, Keivan Stassun, Andreas Berlind, Opie, Mike Line, and others for their extremely constructive feedback. NRH would like to thank CHW3 for his continued support. NRH and GS acknowledge the support of the Vanderbilt Office of the Provost through the Vanderbilt Initiative in Data-intensive Astrophysics (VIDA) fellowship. CTU acknowledges the support of Arizona State University through the SESE Exploration fellowship. RG acknowledges support from the Moore-Sloan Data Science Environment at NYU.
The research shown here acknowledges use of the Hypatia Catalog Database, an online compilation of stellar abundance data as described in \citet{Hinkel14}, which was supported by NASA's Nexus for Solar System Science (NExSS) research coordination network and VIDA.
The results reported herein benefited from collaborations and/or
information exchange within NASA's NExSS research coordination network sponsored by NASA's Science Mission Directorate.
This research has made use of the NASA Exoplanet Archive, which is operated by the California Institute of Technology, under contract with the National Aeronautics and Space Administration under the Exoplanet Exploration Program. 
This research has made use of the services of the ESO Science Archive Facility.
 
\bibliographystyle{aasjournal}
\bibliography{papersML}

\clearpage
\mbox{~}
\clearpage

\widetext
\begin{center}
\textbf{\large Supplemental Materials}
\end{center}
\setcounter{equation}{0}
\setcounter{figure}{0}
\setcounter{table}{0}
\setcounter{section}{0}
\makeatletter
\renewcommand{\theequation}{S\arabic{equation}}
\renewcommand{\thefigure}{S\arabic{figure}}
\renewcommand{\thesection}{S\arabic{section}}

\vspace{-5mm}
\section{Planetary Sample}

For our sample of known planetary host stars, we began with a query to the NASA Exoplanet Archive. We removed 3 planetary systems that were discovered either by direct detection or by the transit method, such that the rest of our sample were all observed via the radial velocity technique. This left us with a total of 316 stars known to host planets. Of those planets, a total of 26 planets had a minimum planet mass below 0.0945 $M_J$ = 30 $M_{\oplus}$, or the mass cutoff between Neptune-like and Jupiter-like planets \citep{Mayor11, Adibekyan12}. 

\begin{figure}[h]
\begin{center}
  \centerline{\includegraphics[width=13cm]{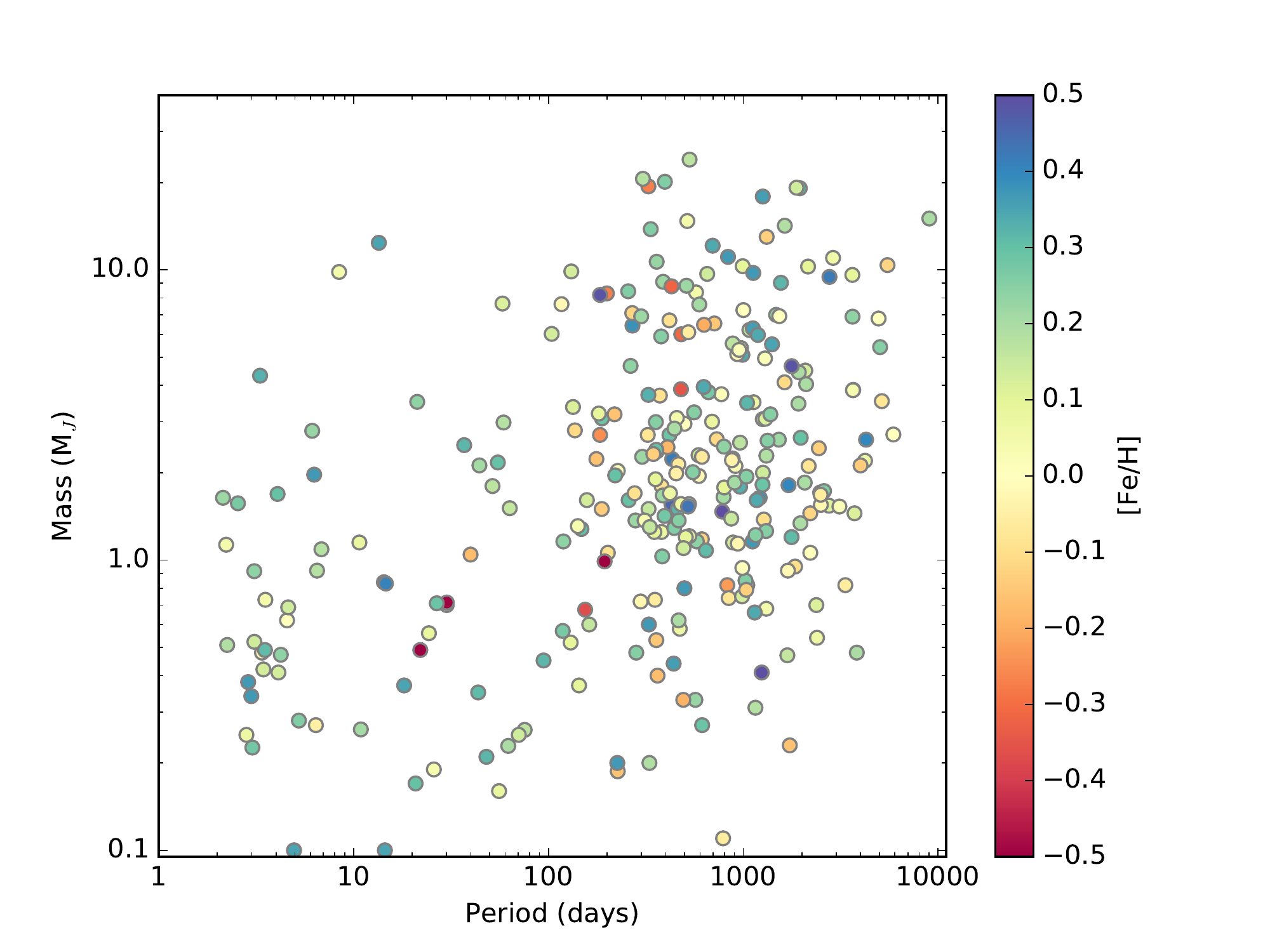}}
\end{center}
  \caption{Log-log plot showing planetary mass versus period for the 290 confirmed planet hosting stars. The stars are color-coded based on the overall [Fe/H] content of the host star such that red is metal-poor, yellow is solar-like, and blue is metal rich.}
  \label{fig:planets}
\end{figure}

We have plotted the 290 confirmed exoplanets by mass and period in Fig. \ref{fig:planets}. There are 30 hot Jupiters, which are defined as any planet with a mass greater than 0.0945 $M_J$ within a 10 day orbital period around the host star, which comprises $\sim$10\% of our training sample. The vast majority of planets (215 planets or 74\%) utilized here have a mass above 0.5 M$_J$ and a period greater than 100 days. The average [Fe/H] content of the confirmed planet hosting stars is 0.13 dex, where 56 host stars (or 20\%) have [Fe/H] $<$ 0.0 dex. This confirms overall the planet-``metallicity" trend originally put forward by \citet{Gonzalez:1997p3950, Santos:2004p2996, Fischer:2005p948} that stars hosting giant planets are preferentially enriched in [Fe/H].

\section{Algorithm Verification}\label{golden}  
Because we know which stars in our sample have been observed to host giant planets, we can evaluate the efficacy of our supervised classifier by running the algorithm on stars with known planets and determining how frequently these planets are correctly predicted. In particular, we can compare the number of true positives (when the model states that a star has a planet when it does have a planet) and false negatives (when a star is not predicted to host a planet but it does have a planet) to assess the frequency of success. We find that our model correctly predicted the existence of a giant exoplanet $\ge$ 90\% of the time for all runs, regardless of the element ensemble. To ensure that we weren't inadvertently biasing our data, we ran the algorithm using a training set where the known and unknown planet host stars were mixed together. This process allowed us to check that when we put in ``noise," we got out ``noise." As a result, we found an equal number of true positives and false negatives, meaning that the algorithm could no longer associate between stars with and stars without giant planets -- as expected.

We also made use of the ``golden sets" when analyzing each of the elemental ensembles, as an additional means to determine if specific element groupings influence the algorithm in different ways. Recall that a ``golden set'' is a set of 10 randomly chosen stars known to host giant planet, which were removed from the training set during each iteration of the algorithm and placed within the target sample. During each run, the algorithm makes a prediction of the likelihood of being a giant planet host (see Sec. 2.4 for more details), for the purpose of testing the algorithm, and then chooses a new golden set. Based on the thousands of golden set predictions, we determined the average prediction likelihood of the known giant exoplanet host stars for each ensemble, as well as the fraction of the ``golden set" that had a probability of hosting a planet which was greater than 90\%. In Section 2.4, we discussed the prediction probabilities arising from the golden set test for the case of Volatiles + Lithophiles + Siderophile + Fe. Listed here are the results from all the ensembles:

\begin{description}
\begin{centering}
\item[Volatiles + Lithophiles + Siderophile + Fe] \hfill \\  Golden Set: Average = 75\%, Above 90\% = 54\%
\item[Volatiles + Lithophiles + Siderophile] \hfill \\  Golden Set: Average = 77\%, Above 90\% = 53\%
\item[Lithophiles + Siderophile + Fe] \hfill \\  Golden Set: Average = 73\%, Above 90\% = 49\%
\item[Volatiles + Lithophiles + Fe] \hfill \\  Golden Set: Average = 76\%, Above 90\% = 58\%
\item[Lithophiles + Siderophile] \hfill \\ Golden Set: Average = 74\%, Above 90\% = 43\%

\end{centering}
\end{description}

\noindent
Overall the results are similar for each ensemble of elements but show an increase in accuracy when Fe and Volatiles are included, as we found when examining feature importance in Section 2.4.

\begin{figure*}
\begin{center}$
\begin{array}{lll}
\includegraphics[trim=0 8mm 0 30mm,clip,width=.4\textwidth]{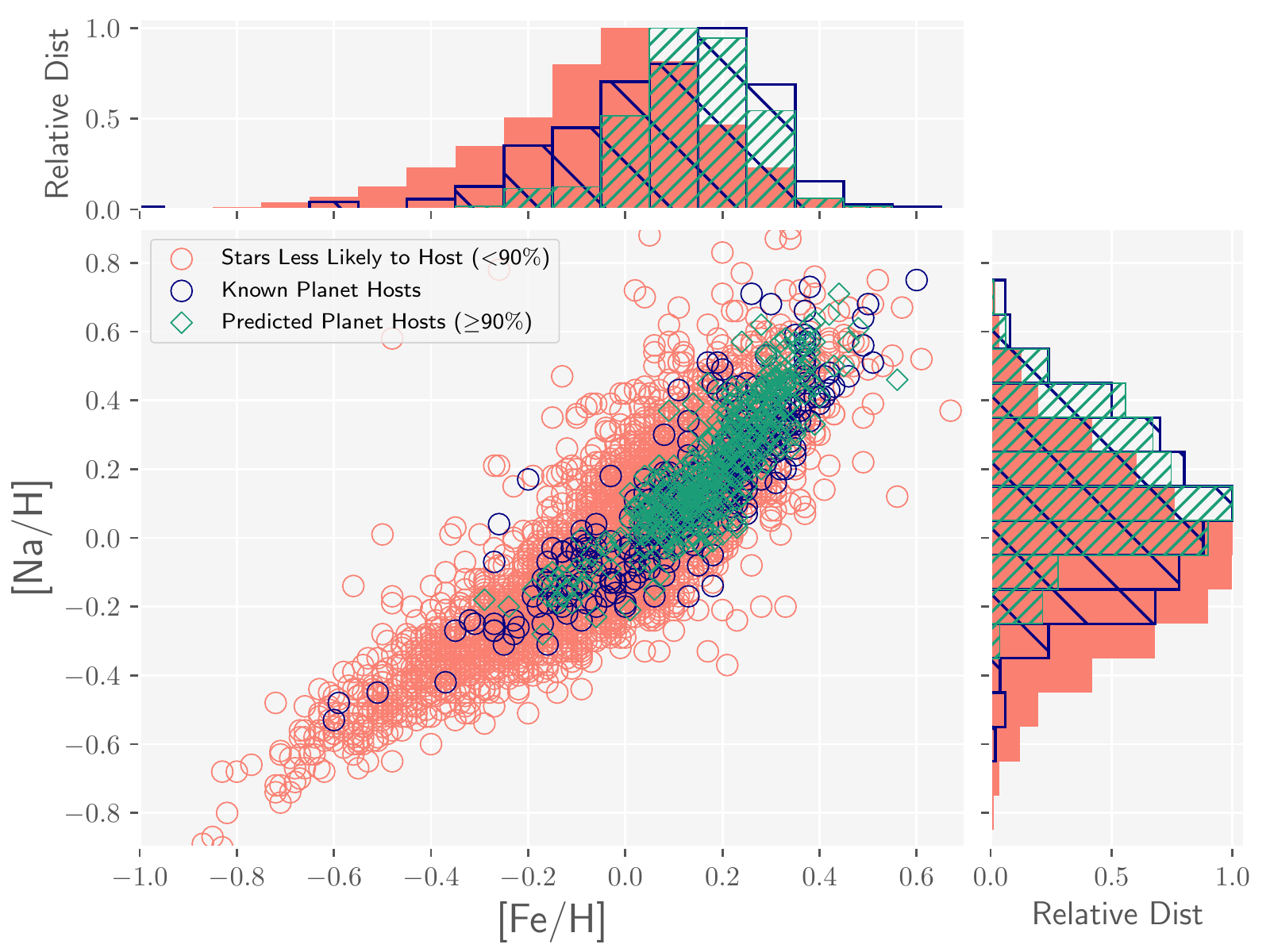} &
\includegraphics[trim=0 8mm 0 30mm,clip,width=.4\textwidth]{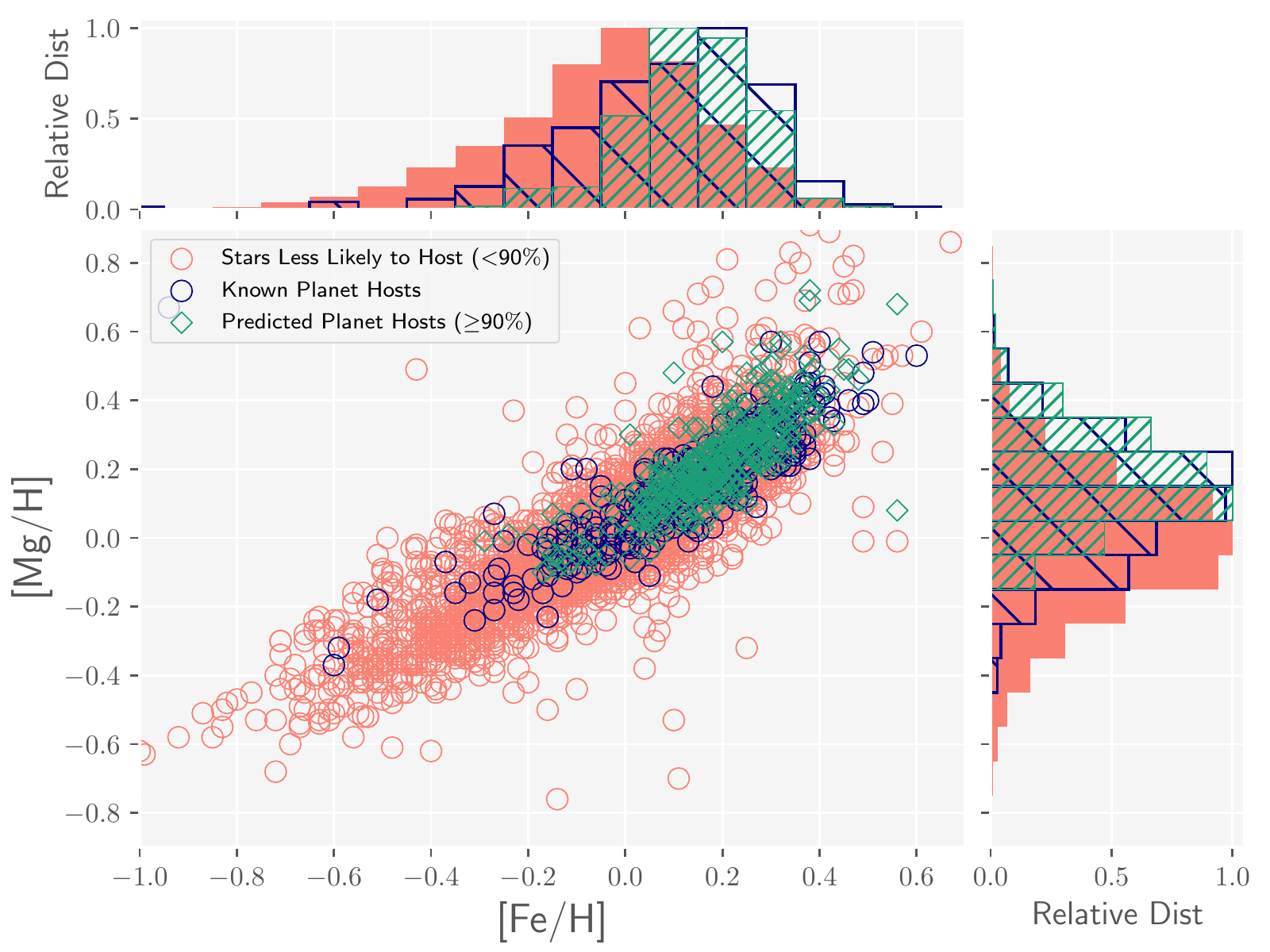}
\end{array}$
\end{center}
\begin{center}$
\begin{array}{rr}
\includegraphics[trim=0 8mm 0 30mm,clip,width=.4\textwidth]{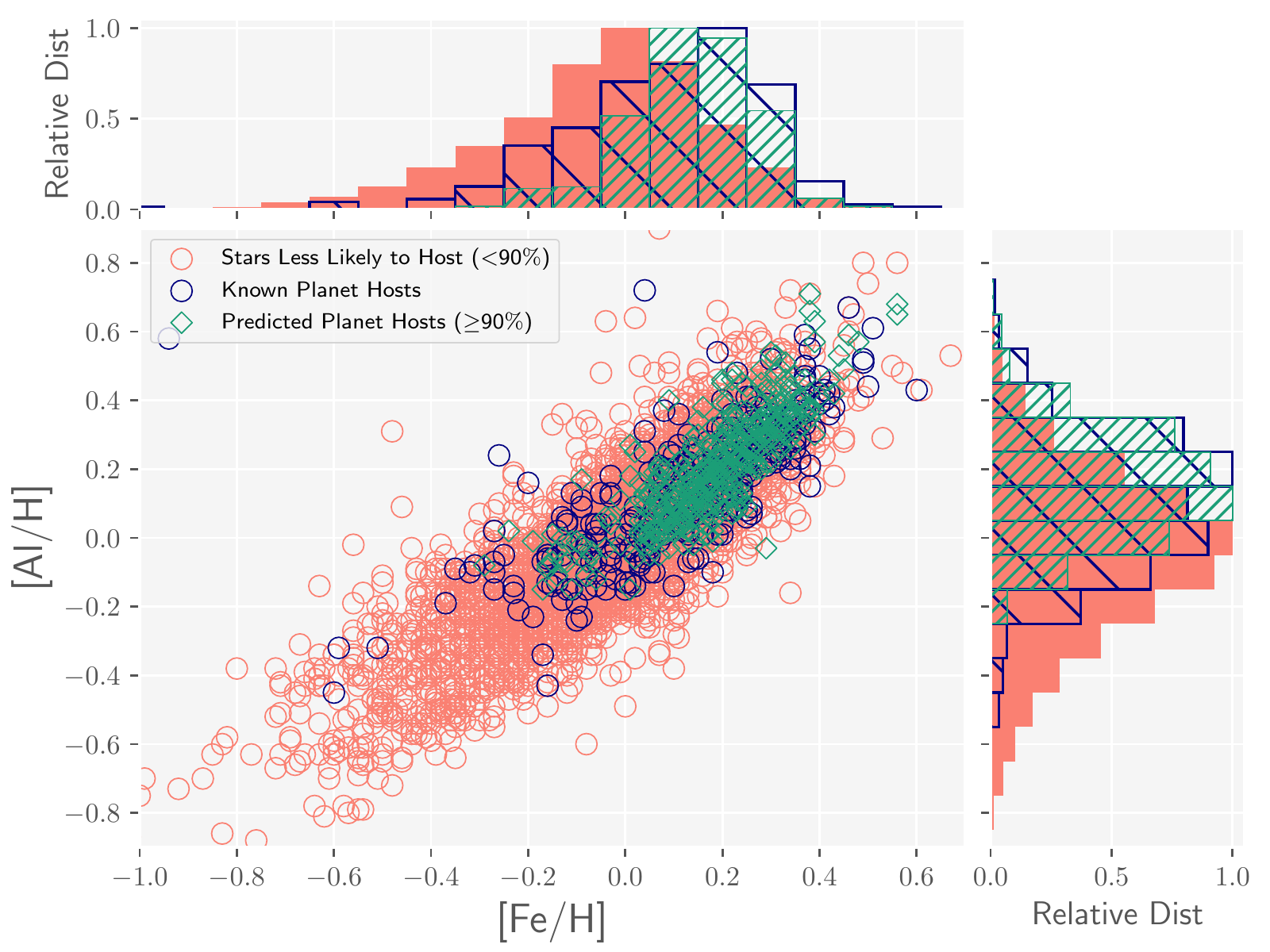} &
\includegraphics[trim=0 8mm 0 30mm,clip,width=.4\textwidth]{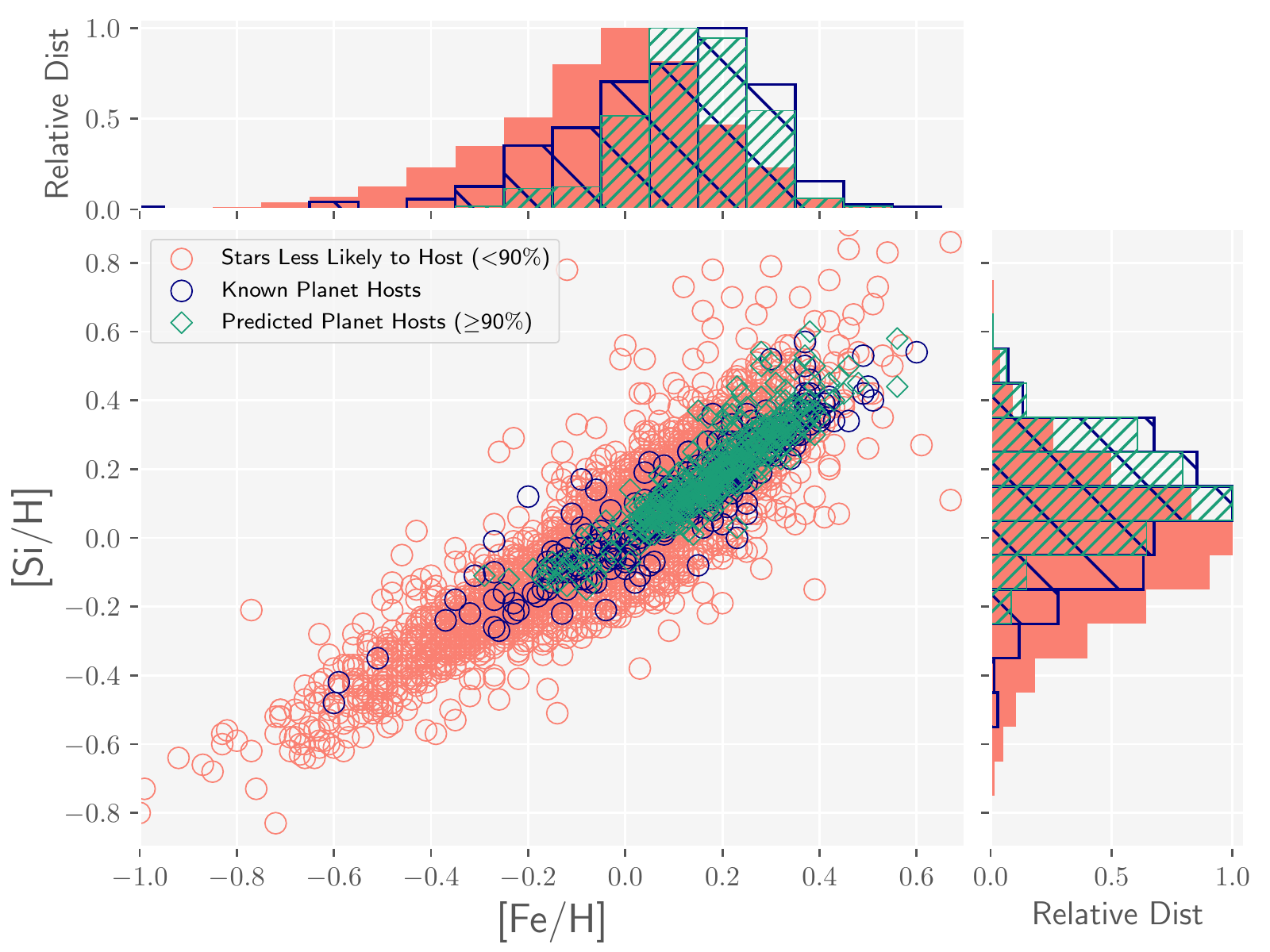} 
\end{array}$
\end{center}
\begin{center}$
\begin{array}{rr}
\includegraphics[trim=0 0 0 30mm,clip,width=.4\textwidth]{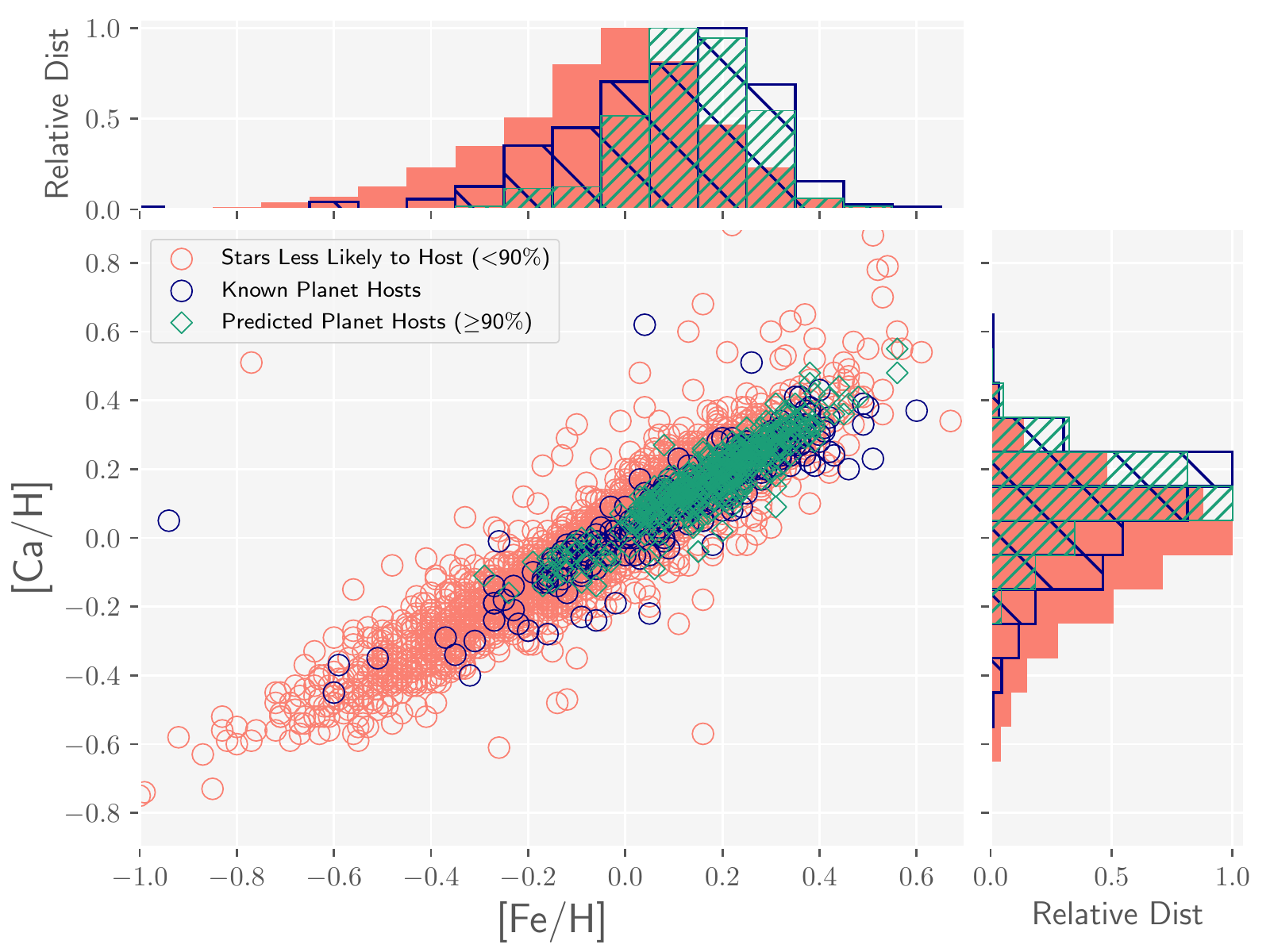} &
\includegraphics[trim=0 0 0 30mm,clip,width=.4\textwidth]{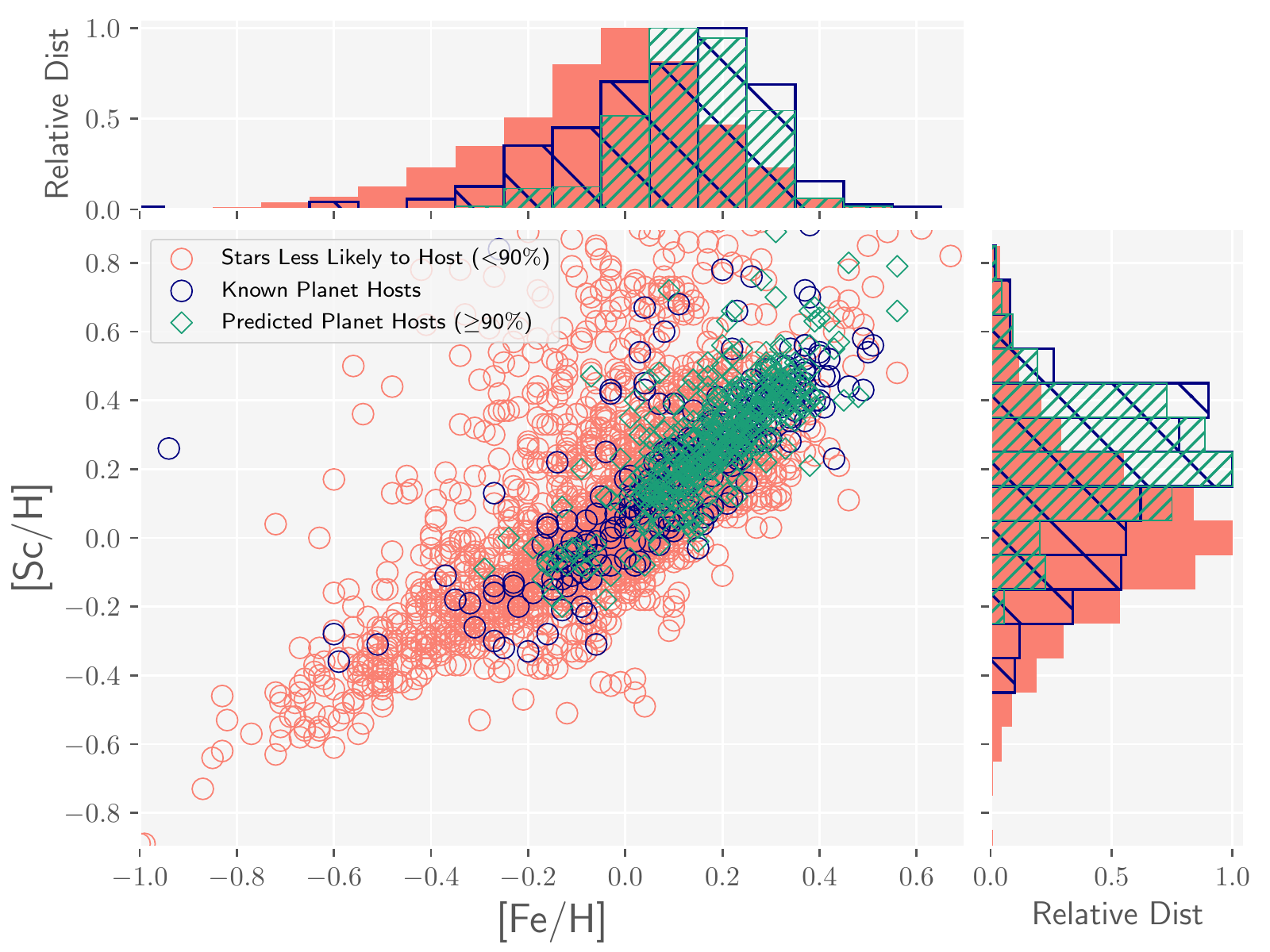} 
\end{array}$
\end{center}
\caption{Multiple [X/H] vs [Fe/H] plots for all of the elements (Vol+Litho+Sidero+Fe, see also Fig \ref{fig:elements2}) where groups of stars are color-coded to indicate stars less likely to host giant planets (orange), known planet host stars (navy), and stars that were given a $\ge$90\% prediction score via the algorithm (green). One-dimensional histograms of the relative [X/H] distribution are shown to the right of all scatterplots while the [Fe/H] is shown at the top.}
\label{fig:elements1}
\end{figure*}

\begin{figure*}
\begin{center}$
\begin{array}{lll}
\includegraphics[trim=0 8mm 0 0,clip,width=.4\textwidth]{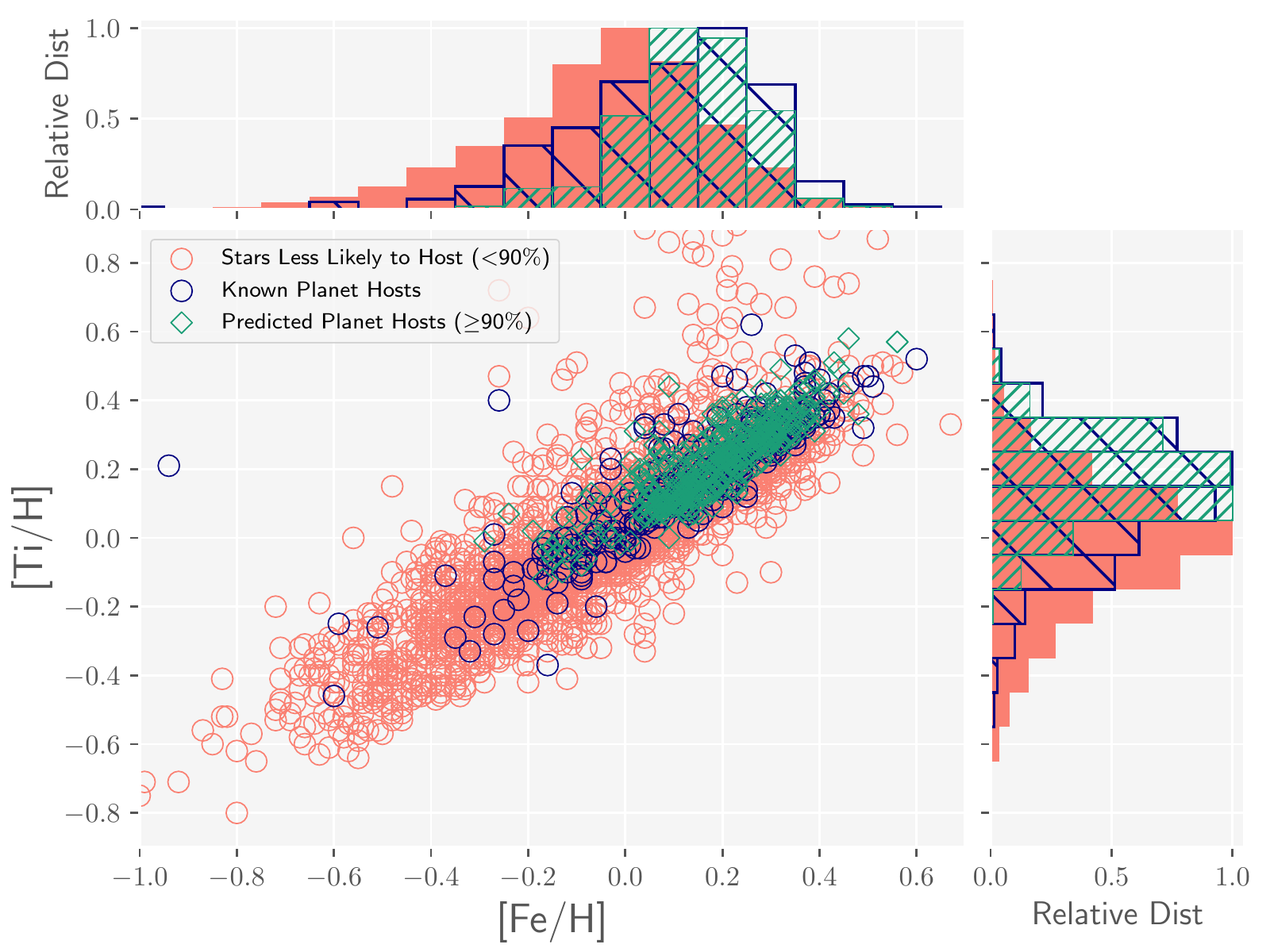}&
\includegraphics[trim=0 8mm 0 0,clip,width=.4\textwidth]{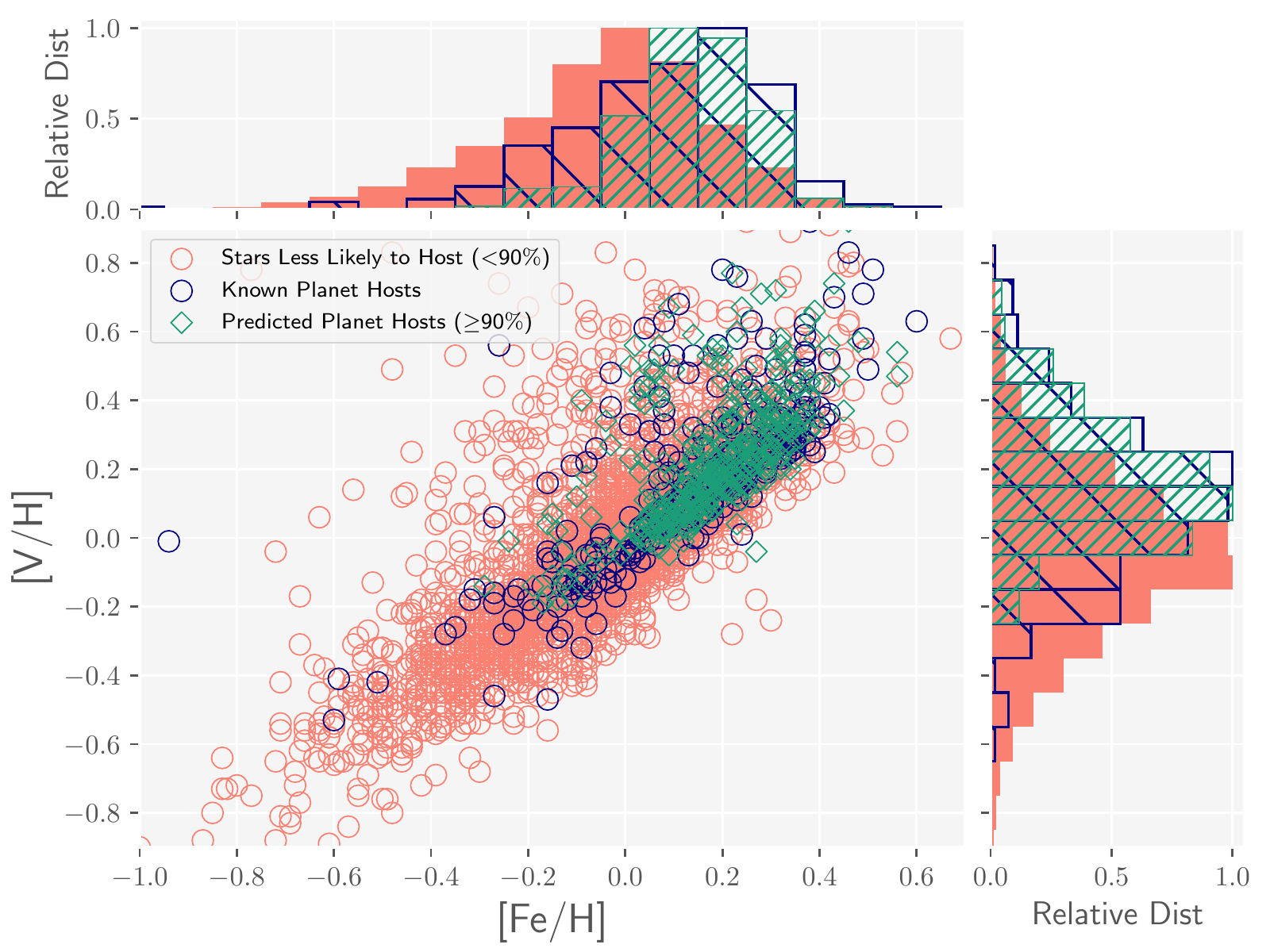}
\end{array}$
\end{center}
\begin{center}$
\begin{array}{rr}
\includegraphics[trim=0 8mm 0 30mm,clip,width=.4\textwidth]{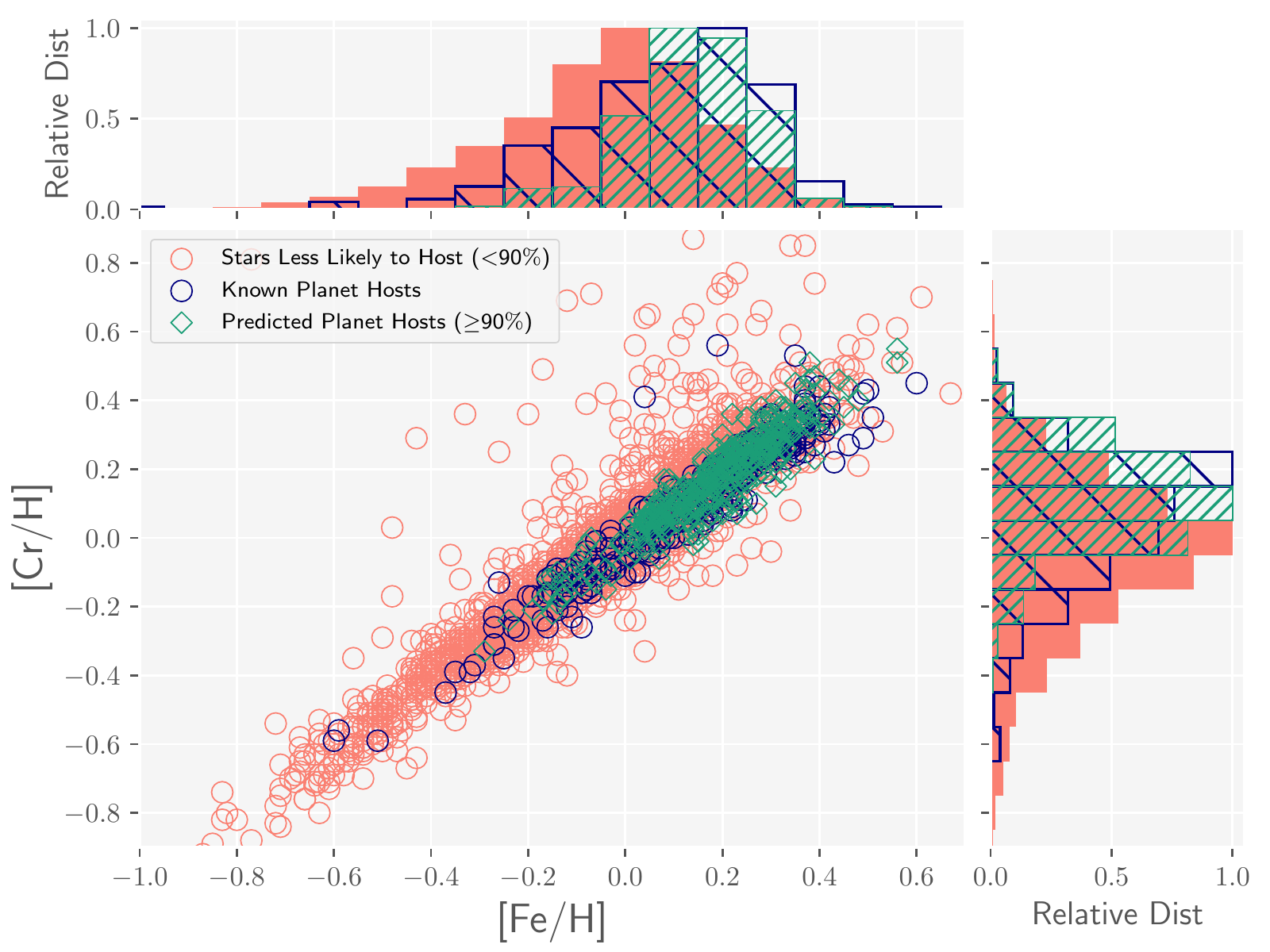} &
\includegraphics[trim=0 8mm 0 30mm,clip,width=.4\textwidth]{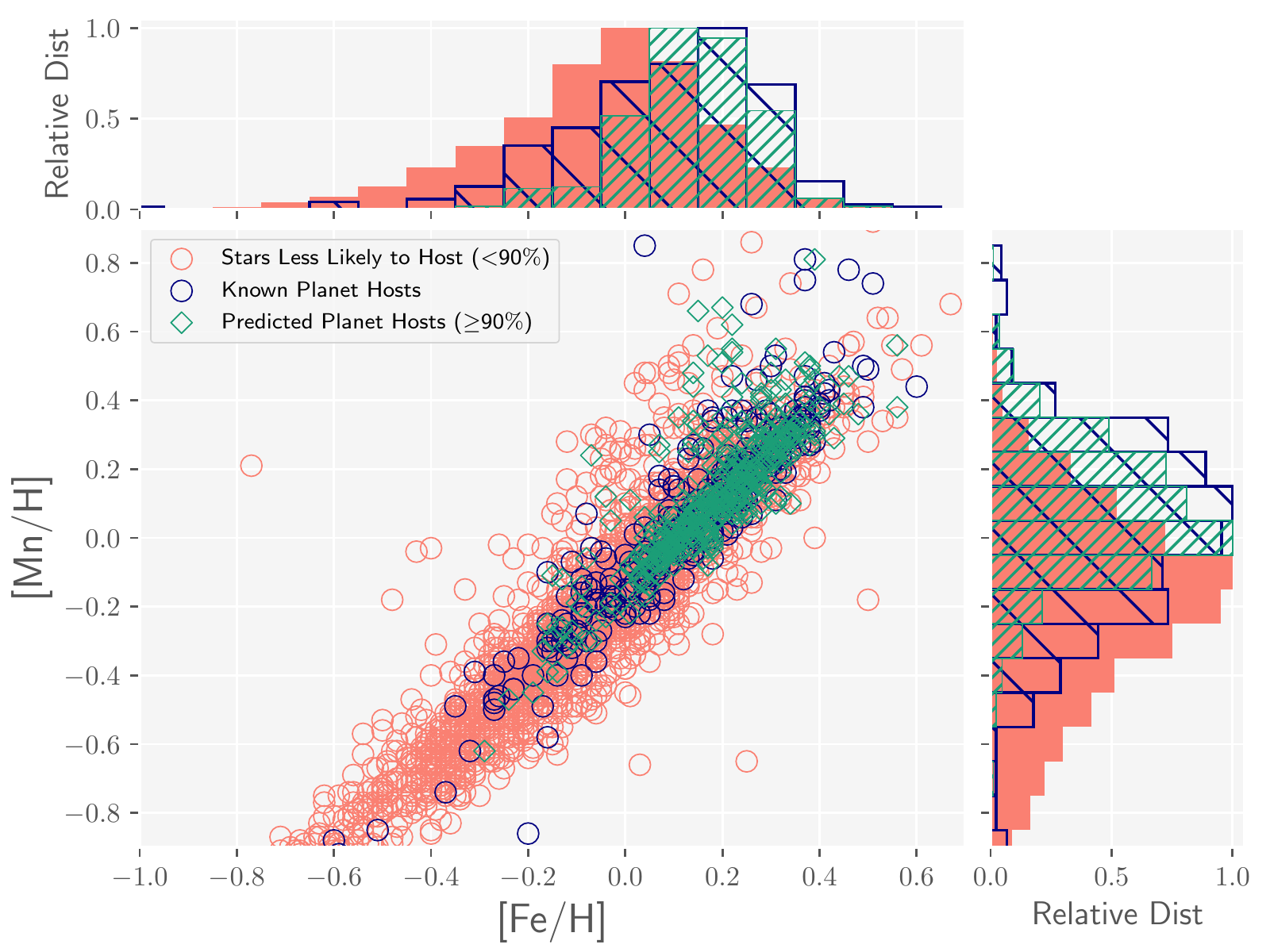}
\end{array}$
\end{center}
\begin{center}$
\begin{array}{rr}
\includegraphics[trim=0 0 0 30mm,clip,width=.4\textwidth]{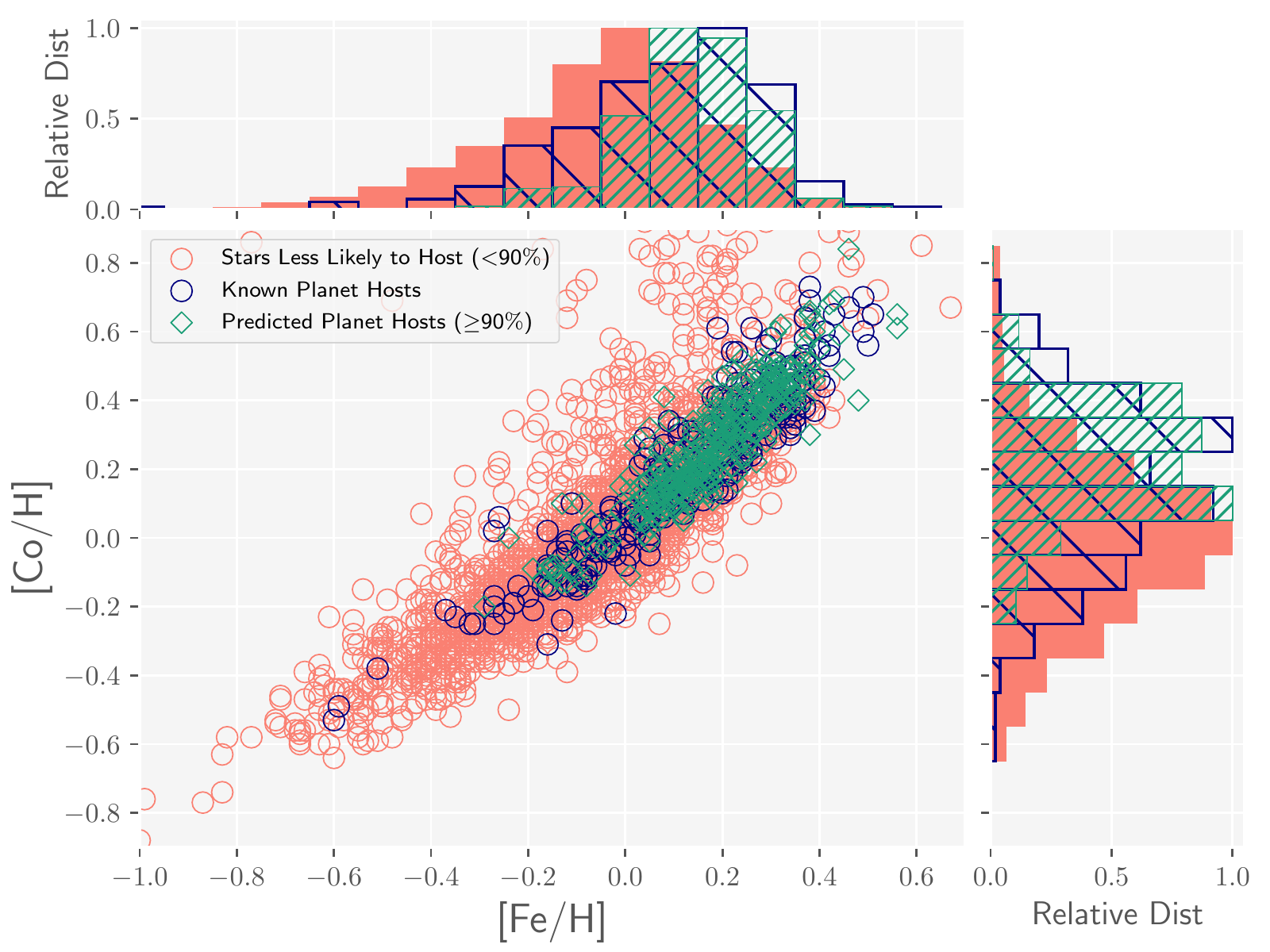} &
\includegraphics[trim=0 0 0 30mm,clip,width=.4\textwidth]{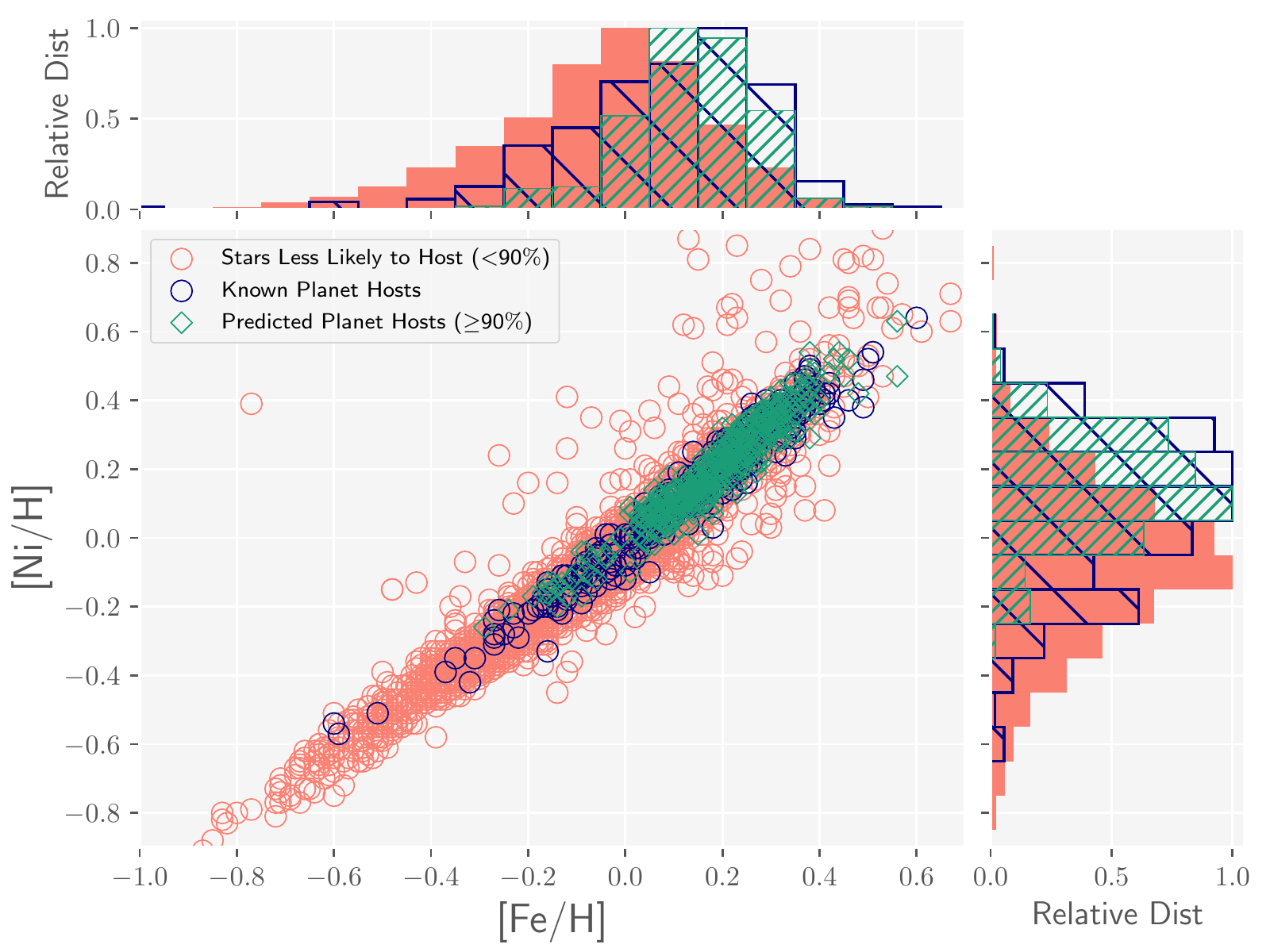} 
\end{array}$
\end{center}
\begin{center}$
\begin{array}{rr}
\includegraphics[trim=0 0 0 30mm,clip,width=.4\textwidth]{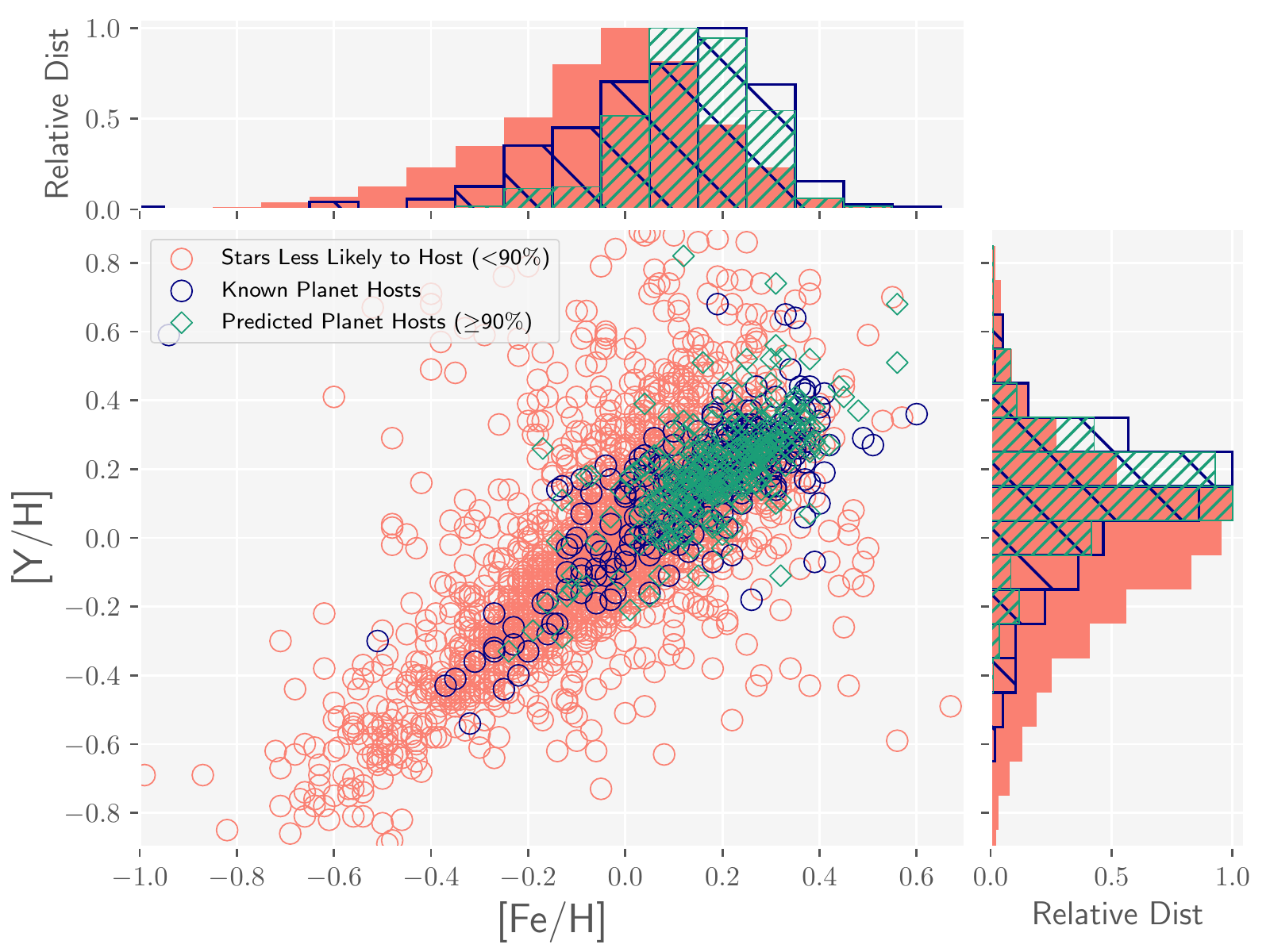} 
\end{array}$
\end{center}
\caption{Continuation of Fig. \ref{fig:elements1}.}
\label{fig:elements2}
\end{figure*}

\section{Variation Between Ensembles}\label{ens-variations}

The feature importance score is calculated individually for each feature within the dataset, in this case, the elements we included in each ensemble. Their importance is determined for each single decision tree -- there are 1000 decision trees per iteration, 3000 iterations per run, we performed hundreds of runs across the five ensembles. The score is calculated by the amount that each feature's split point -- which is proposed by the algorithm based on the percentiles of feature distribution -- improves the performance measure. In this case, performance is measured towards being a likely giant exoplanet host star, based on the training set. The value is then weighted by the number of observations per leaf node in the tree and then averaged across all decision trees within the model. The discrete calculations are outlined in \citet{Chen16}.

When beginning with a large dataset, the features of the data can be described as being somewhat random and, as a result, the dataset can be considered to have high entropy. 
It is by splitting the larger dataset into smaller ones, with similar properties, that the overall entropy can be lowered until you have groupings of like-stars. The usefulness of a property, or feature, in making a decision (for example, proper motion would be more useful than parallax in determining whether a star originated from the thin or thick disk) is indicated by its feature importance score. The more a feature is used to make pivotal choices within the decision tree that ultimately lower the entropy, the higher its importance score. 

For this paper, the important features are those elements that are influential in splitting the overall sample into stars likely to host a giant exoplanet or not. As mentioned in Section \ref{ensembles}, we examine multiple ensembles of elements in order to better understand the overall correlation of their presence on each other and on their decisions. With that in mind, we have two caveats when determining the feature importance scores: (1) When present, Fe must have a high feature importance score -- since its influence on giant planet detections has been consistently seen in observations; and (2) The presence of a null element abundance measurement for a star should not impact the decision tree. While the latter point seems obvious, the assumption within {\it XGBoost} is that the decision goes in the ``default" direction of the training set, in this case, towards the population of stars predicted to host a giant exoplanet. It was therefore necessary to include a weighting function which would remove the impact of the null values on the feature importance score without overcorrecting the problem. This was done by testing the weighting function on more complete datasets i.e. with only a small number of elements that were more commonly measured in nearby stars and which had the fewest number of null values. We then slowly included less populated data (i.e. less frequently measured elements) and evaluated the weighting function. We continued propogating the weighting function until we saw noticeable change in the behavior of the output which signified an overcorrection, such that a preference was given towards those elements with the most number of null values
\footnote{The behavior of the null features, and the desire to avoid overcorrecting the problem, resulted in us including fewer elements within our ensembles as originally hoped.}. Finally, the features were normalized such that the highest feature always had a score of 1.0, to make it easier to compare scores between different ensembles.

The sampling results for all five ensembles are listed in Supplement table. This table is ordered and sorted with preference towards those ensembles which contained a higher number of elements, since the results of the ensembles with fewer elements were mirrored in those with more features (per Figure \ref{fig:importance}). Similar to Table 1, the RA, Dec, spectral type, and V magnitude are included for all of the stars to make it as useful as possible for follow-up observations. Additionally, the raw number of times a star was sampled and then predicted to have a planet are included for transparency.

\section{Closer Examination of Vol+Litho+Sidero+Fe}
We focus now on exploring the specific results produced after running the prediction algorithm for a single ensemble of elements, namely, the Vol+Litho+Sidero+Fe ensemble which contained all of the elements utilized in this paper. Having shown the weighted feature importance score for this ensemble at the bottom of Fig. \ref{fig:importance}, we have also plotted the individual [X/H] vs [Fe/H] plots for all elements as illustrated in Figs. \ref{fig:elements1} and \ref{fig:elements2}. In each subplot, we show the training sample of the 290 known planet host stars in navy. The target sample of the +4200 stars not known to host planets are broken into two groups: those stars with a $\ge$90\% probability of hosting a giant planet in green and stars with a less likely probability ($<$90\%) of hosting a giant planet in orange (see Section \ref{predict} for more discussion). Each of the scatter plots has a corresponding [X/H] relative frequency histogram located on the right, while the [Fe/H] histogram (which is the same for all scatter plots) is seen at the top of both columns in Figs. \ref{fig:elements1} and \ref{fig:elements2}. The bins of all histograms have a width of 0.1 dex. 

When analyzing the scatter plots in Figs. \ref{fig:elements1} and \ref{fig:elements2}, it is not our intention to imply that planet- and predicted-planet-hosting stars are all enriched with respect to these elements. Instead, our purpose is to compare the overall distributions of the stellar populations, especially between the known and predicted planet host stars. We see that the known and predicted planet host stars (navy and green, respectively) lie within the same region of parameter space for all of the elements within this ensemble, to within typical, respective error for each element \citep{Hinkel14}. The strong overlap offers a visual confirmation that the trends in stars predicted to host planets match the trends of stars known to host giant planets, even when analyzed on an element-by-element basis. The same cannot be said when comparing the stars less likely to host a planet (orange) to the other two populations. In terms of [Fe/H] content, the stars without a high ($\ge$90\%) probability of hosting a planet have, in general, a lower [Fe/H] content than stars known or predicted to host giant exoplanets -- a variation which is greater than the typical $\pm$0.05 dex error for [Fe/H] as confirmed in literature. We also see a strong variation between the less likely to host population and the known/predicted planet hosts when looking at [C/H], [Sc/H], and [Ni/H] -- as noted by the difference of at least two bin widths between the populations in the histograms. For C, the differences were important in deciding which stars were likely to host giant exoplanets, as discussed in Section \ref{ens-variations}. However, for Sc and Ni the dispersions between the populations may have made some difference in determining which stars were predicted to host planets, although the two elements were often in the lower half of the feature importance scores, see particularly Vol+Litho+Sidero+Fe. Additionally, there is a somewhat bimodal trend in [Co/H] between the stars known to host planets and the stars predicted to have planets, indicated by the histogram on the y-axis, which likely resulted in its having the lowest or second lowest (for Vol+Litho+Sidero+Fe) feature importance score in every instance where it was included within the ensemble (see Fig. \ref{fig:importance}).

\clearpage
\begin{turnpage}
\begin{deluxetable}{p{0.7cm}p{0.4cm}p{1.0cm}p{1.2cm}p{0.2cm}|p{0.4cm}p{0.4cm}p{0.1cm}|p{0.4cm}p{0.4cm}p{0.4cm}|p{0.4cm}p{0.4cm}p{0.4cm}|p{0.4cm}p{0.4cm}p{0.4cm}|p{0.4cm}p{0.4cm}p{0.4cm}} 
  \tablecaption{\label{predtable2}Predicted Giant Exoplanet Host Stars}
  \tablehead{
\colhead{HIP} &  \colhead{RA} & \colhead{Dec } & \colhead{Spec} &  \colhead{ V  }  & \multicolumn{3}{|c|}{Vol+Litho+Sidero+Fe} & \multicolumn{3}{c|}{Vol+Litho+Sidero} & \multicolumn{3}{c|}{Vol+Litho+Fe} & \multicolumn{3}{c|}{Litho+Sidero+Fe} & \multicolumn{3}{c}{Litho+Sidero}
\\
\colhead{ } &  \colhead{(deg)} & \colhead{(deg)} & \colhead{ Type } &   \multicolumn{1}{c|}{  mag } & \colhead{Samp} & \colhead{ Pred} &   \multicolumn{1}{c|}{Prob } & \colhead{Samp} & \colhead{ Pred} &  \multicolumn{1}{c|}{ Prob } & \colhead{Samp} & \colhead{ Pred} &   \multicolumn{1}{c|}{ Prob } & \colhead{Samp} & \colhead{ Pred} &  \multicolumn{1}{c|}{ Prob } & \colhead{Samp} & \colhead{ Pred} &  \colhead{ Prob } 
}
62345 & 191.63 & -11.81 & G5V & 6.87 & 2882 & 2882 & 1.0 & 2873 & 2873 & 1.0 & 2860 & 2860 & 1.0 & 2881 & 2881 & 1.0 & 2864 & 2864 & 1.0 \\ 
24110 & 77.68 & -44.57 & G8IV/V & 8.71 & 2878 & 2878 & 1.0 & 2868 & 2868 & 1.0 & 2858 & 2858 & 1.0 & 2887 & 2884 & 0.999 & 2869 & 2866 & 0.999 \\ 
111978 & 340.23 & -31.99 & K0IV-V & 7.39 & 2856 & 2856 & 1.0 & 2870 & 2870 & 1.0 & 2866 & 2866 & 1.0 & 2880 & 2879 & 0.999 & 2882 & 2881 & 0.999 \\ 
20489 & 65.87 & -27.66 & G3/5V & 8.58 & 2863 & 2863 & 1.0 & 2873 & 2873 & 1.0 & 2868 & 2867 & 0.999 & 2865 & 2863 & 0.999 & 2879 & 2879 & 1.0 \\ 
81347 & 249.20 & -6.29 & G5V & 7.82 & 2861 & 2861 & 1.0 & 2864 & 2863 & 0.999 & 2861 & 2855 & 0.998 & 2870 & 2870 & 1.0 & 2871 & 2871 & 1.0 \\ 
68936 & 211.67 & -5.52 & K1V & 8.36 & 2844 & 2844 & 1.0 & 2855 & 2853 & 0.999 & 2865 & 2863 & 0.999 & 2862 & 2831 & 0.989 & 2881 & 2865 & 0.994 \\ 
116823 & 355.17 & 0.42 & K2III & 7.44 & 2846 & 2846 & 1.0 & 2885 & 2882 & 0.999 & 2891 & 2891 & 1.0 & 2875 & 2872 & 0.999 & 2860 & 2856 & 0.999 \\ 
71803 & 220.32 & -4.94 & G6V & 8.21 & 2877 & 2877 & 1.0 & 2883 & 2880 & 0.999 & 2844 & 2841 & 0.999 & 2870 & 2864 & 0.998 & 2861 & 2853 & 0.997 \\
... & ... & ... & ... & ... & ... & ... & ... & ... & ... & ... & ... & ... & ... & ... & ... & ... & ... & ... & ... 
\enddata
 \tablenotetext{*}{This is a stub of the table, available via the online journal and Vizier, for the full sample of target stars that were predicted on. The table lists the HIP name, RA, Dec, spectral type, V magnitude, as well as the number of times that the star was sampled (Samp), predicted to host a giant planet (Pred), and the overall probability (Pred/Samp) of hosting a giant planet for all 5 ensembles of elements -- which are listed from most number of elements to fewest.}
\end{deluxetable}
\clearpage
\end{turnpage}

\end{document}